\documentclass{ametsocV6.1}
\nolinenumbers
\let\linenumbers\relax
\let\internallinenumbers\relax

\usepackage{hyperref}
\usepackage[inline]{trackchanges} 
\addeditor{AGF}
\addeditor{MEG}
\addeditor{JCP}

\emergencystretch=\maxdimen
\title{Assessing Diffusion and Rectified Flow Models for Multivariate Statistical Climate Downscaling in the Canary Islands} 

\authors{Marcos Esquivel-González,\aff{a}\correspondingauthor{Marcos Esquivel-González, fgmesquivel@ull.edu.es} 
Albano González,\aff{a} 
Juan Carlos Pérez,\aff{a} 
Jose González-Abad,\aff{b} 
and Juan Pedro Díaz\aff{b} 
}

\affiliation{\aff{a}{Universidad de La Laguna (ULL), Spain}\\
\aff{b}{Instituto de Física de Cantabria (IFCA), CSIC-Universidad de Cantabria, Spain}
}

\nolinenumbers
\abstract{
As climate-impact studies increasingly require local-scale information, statistical downscaling provides a computationally efficient way to derive high-resolution climate fields from coarse-resolution global simulations. However, regression-based deep learning methods often produce overly smooth precipitation fields, provide limited information on predictive uncertainty, and struggle to represent extremes. This study evaluates latent generative models for multivariate downscaling over the complex domain of the Canary Islands using ERA5 predictors and high-resolution ROCIO+\_CAN observations of daily maximum temperature, minimum temperature, and precipitation for 1990--2022. A Gaussian diffusion model, ClimDiT, and a rectified flow variant, ClimDiT-RF, together with their 50-member ensembles, are compared against deterministic and stochastic baselines. Methodological contributions include a lighter ClimDiT backbone, a reduced 100-step denoising diffusion sampling configuration, and an extreme-aware composite validation score for checkpoint selection. A key finding is that RMSE-only selection can favor checkpoints with lower average errors but degraded extreme percentiles, whereas the composite criterion provides a better balance and improves selection for generative and baseline models as well. ClimDiT achieves competitive temperature downscaling comparable to the best deterministic baselines and improves the spatial representation of precipitation, particularly its spectral power and intense precipitation. The generative models also achieve the lowest deterministic multivariate errors, while their ensembles improve predictive uncertainty estimates, although probabilistic gains for precipitation remain modest. ClimDiT-RF provides competitive deterministic performance but tends to underestimate precipitation variability, spectral power, and extremes more strongly than Gaussian diffusion. Overall, latent Gaussian diffusion is a promising framework for multivariate climate downscaling, given its evidenced effectiveness in complex, high-resolution domains.
} 

\begin{document}

\maketitle

%
%
%
\statement
Reliable local climate information is especially difficult to produce for regions such as the Canary Islands, where steep terrain creates sharp changes in temperature and rainfall over short distances. We tested new computer-based methods designed to generate detailed climate maps from coarse atmospheric data. The best-performing approach reproduced local temperature patterns accurately and represented fine-scale rainfall structure and heavy precipitation more realistically than several widely used alternatives. It also produced multiple plausible outcomes, providing useful information about uncertainty at a manageable computational cost. We further found that selecting a model checkpoint during training from average errors alone can hide a worsening representation of extremes; accounting for both average and extreme-event errors during validation produced a better balance. These findings support the development of more reliable local climate datasets for impact studies and adaptation planning in mountainous and island regions. Future work should test the method in other climates and under future warming conditions.
%
%

%
\section{Introduction}

High-resolution climate data are essential for assessing the localized impacts of climate variability and climate change, particularly in regions characterized by complex topography. Dynamical downscaling using regional climate models (RCMs) can provide physically consistent regional simulations, but its high computational cost limits the generation of large ensembles and the systematic exploration of uncertainty. Statistical downscaling (SD) therefore provides a complementary and widely used alternative by learning empirical relationships between coarse-resolution atmospheric predictors and high-resolution local predictands \citep{maraun2010precipitation}.

Deep learning (DL) methods, particularly convolutional neural networks (CNNs), have improved the performance of SD systems relative to many traditional approaches \citep{bano2020deepesd,vandal2017deepesd}. However, when neural networks are trained with deterministic loss functions such as mean squared error, they usually approximate conditional means. This can produce overly smooth fields and contribute to the underestimation of spatial variability, intermittency, and extreme values \citep{bano2020deepesd,GonzalezAbad2025,ledig2017photo,maraun2010precipitation,vandal2017deepesd,wang2021deep}. 

Generative models provide a promising route to mitigate this limitation because they can represent predictive distributions conditioned on the same large-scale atmospheric state, rather than a single deterministic estimate \citep{rampal2026cordexmlbench}. One relevant class is flow matching, exemplified by \cite{wetherell2026}, which uses a U-Net backbone trained to transport samples from a Gaussian source to the target distribution via a linear scheduler. Generative adversarial networks constitute another widely used family of models. For instance, ResGAN combines a deterministic estimate of the conditional mean with an adversarial model that generates stochastic residuals, thereby improving the representation of precipitation extremes and small-scale spatial structures \citep{rampal2025reliable}. Similarly, SpaGAN uses a convolutional generator and discriminator, together with stochastic noise conditioning, to produce spatially coherent precipitation realizations \citep{glawion2023spategan}. 

Diffusion-based approaches have also become increasingly prominent. CorrDiff learns the residuals in a similar approach to ResGAN, but using a diffusion process \citep{mardani2025}. In turn, RCMGEM \citep{addison2026} relies on score-based generative modelling formulated in terms of stochastic differential equations to generate multivariate regional climate fields conditioned on large-scale atmospheric predictors and high-resolution orography.  
Latent diffusion models (LDMs) have also shown considerable potential for generating spatially coherent and multivariately consistent high-resolution climate fields \citep{EsquivelGonzalez2026climdit,leinonen2023latent}. Specifically, ClimDiT \citep{EsquivelGonzalez2026climdit} combines a latent representation of the target variables with a diffusion transformer, allowing fine-scale structures and intervariable dependencies to be modelled more explicitly.

This study evaluates latent generative models for the multivariate statistical downscaling of daily maximum temperature, daily minimum temperature, and precipitation over the Canary Islands  at a high spatial resolution of 2.5 km. This volcanic archipelago is characterized by highly complex topography, which strongly influences the spatial distribution of these meteorological variables \citep{cropper2014macaronesia}. These properties make these islands a particularly interesting region in which to analyze the performance of SD models, given the associated difficulties. An improved version of the Gaussian diffusion model ClimDiT is evaluated over this region, building on the promising results obtained by the original model in a previous study conducted at coarser resolution over the Iberian Peninsula, a domain characterized by comparatively simpler topography \citep{EsquivelGonzalez2026climdit}. Furthermore, a variant based on an optimized flow matching technique, known as rectified flow, is proposed. Rectified flow ensures that transport trajectories are as straight as possible, which substantially improves computational efficiency and sampling speed \citep{liu2022flow}. The results of both generative models are compared with those obtained using several deterministic and probabilistic CNN-based statistical downscaling models. These baseline models were chosen because they belong to the family of methods that form the basis of the datasets generated under the Spanish National Climate Change Adaptation Plan, PNACC \citep{CorreaGuinea2023,hernanz_lazaro_2026_20338489}. Particular attention is paid to checkpoint selection. Specifically, the study examines whether validation RMSE alone may mask a deterioration in tail behaviour during training and introduces a composite score that jointly monitors mean errors and extreme-percentile errors. Applying this composite criterion to both ClimDiT and the multivariate CNN baselines allows us to assess whether a more comprehensive checkpoint-selection strategy can improve downscaling performance across different model families. Finally, the evaluation includes deterministic, probabilistic, spatial, spectral, and multivariate metrics, together with an extensive analysis of model components, sampling configurations, and computational costs.

The remainder of this paper is structured as follows. Section \ref{sec:Methods} describes the data and methodological framework, including the benchmark models and evaluation metrics. Section \ref{sec:Results} presents and discusses the results. Section \ref{sec:Conclusions} summarizes the main findings, limitations, and broader implications.

\section{Data and methodology}\label{sec:Methods}

\subsection{Study area and data}
\begin{figure}[htbp]
    \centerline{
    \includegraphics[width=0.9\textwidth]{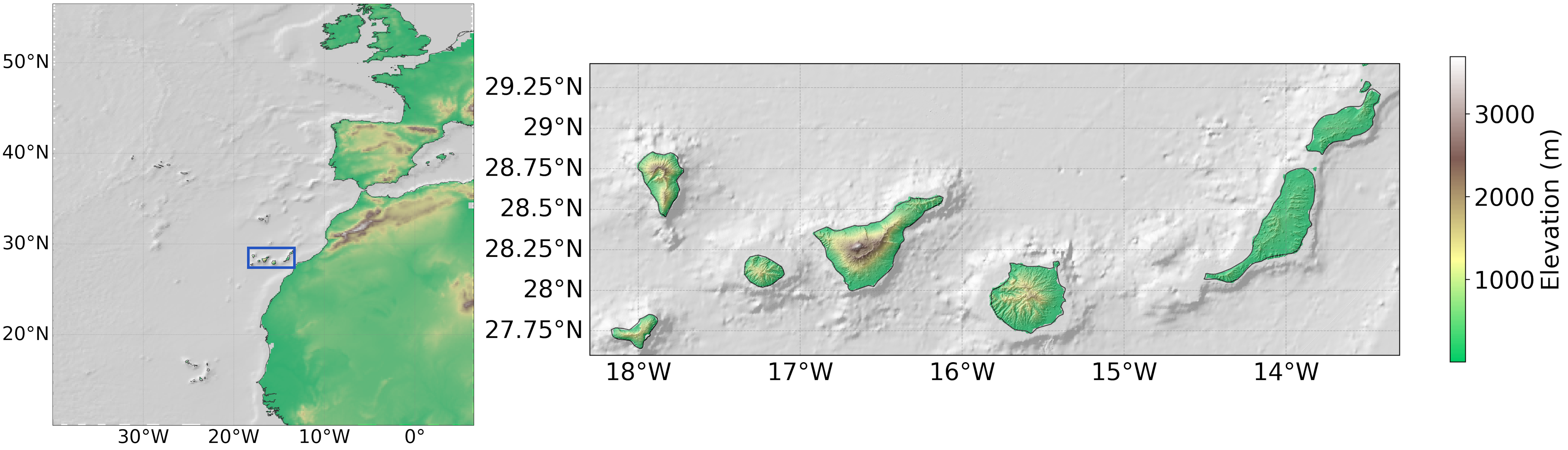}}
    \caption{The left panel shows the ERA5 domain, while the right panel shows the high-resolution domain over which the downscaling is performed. In both panels, terrain elevation (m) is displayed to illustrate the topographic complexity of the study area, with shaded relief used to emphasize the main topographic features. The blue rectangle indicates the location of the Canary Islands domain within the low-resolution ERA5 domain.}
    \label{fig:domain}
\end{figure}

The Canary Islands constitute a particularly challenging testbed for statistical climate downscaling, due to their complex topography and marked spatial climate variability. The interaction between large-scale atmospheric conditions and local topographical features generates marked contrasts in temperature and precipitation, making the archipelago particularly appropriate for evaluating the ability of downscaling models to reproduce climate patterns at a fine scale. The archipelago exhibits a marked west–east topographic contrast: the western islands are steep and mountainous, reaching exceptionally high elevations (3,715 m at Mount Teide) despite their limited surface area, whereas the eastern islands are generally lower and much flatter (Fig. \ref{fig:domain}). This pronounced topographic heterogeneity contributes to the mentioned spatial variations in local climate \citep{LUQUESOLLHEIM2024}. 

Following the setup adopted in previous studies \citep{EsquivelGonzalez2026climdit, AreDeepLearningMeth}, large-scale atmospheric predictors are extracted from the ERA5 reanalysis \citep{Hersbach2020}, while high-resolution local predictands are obtained from a regional gridded observational dataset, in this case ROCIO+\_CAN \citep{ROCIO-IBEB}. ROCIO+\_CAN comprises daily maximum temperature (Tmax), daily minimum temperature (Tmin) and daily accumulated precipitation (Prec) fields at a spatial resolution of 2.5 km. The ERA5 predictors have been conservatively remapped to 1.5-degree grid, which is representative of the typical spatial resolution of global climate models, because the purpose of developing regionalization methods is their application to climate projections generated by such global models. Table \ref{tab:Variables} lists the ERA5 predictors used and the ROCIO+\_CAN target variables, while Figure \ref{fig:domain} represents the spatial domains covered by both datasets.

\begin{table}
\begin{center}
\begin{tabular}{lll}
\topline
\textbf{Source} & \textbf{Variable} & \textbf{Units / Levels} \\
\midline
\textbf{ERA5} & Air temperature & K (850, 700, and 500 hPa) \\
(Predictors) & Specific humidity & kg kg$^{-1}$ (850, 700, and 500 hPa) \\
& Meridional wind ($v$) & m s$^{-1}$ (850, 700, and 500 hPa) \\
& Zonal wind ($u$) & m s$^{-1}$ (850, 700, and 500 hPa) \\
& Mean sea level pressure & Pa \\
\midline
\textbf{ROCIO+\_CAN} & Daily maximum temperature (Tmax) & K \\
(Target variables) & Daily minimum temperature (Tmin) & K \\
& Daily accumulated precipitation (Prec) & mm \\
\botline
\end{tabular}
\caption{ERA5 atmospheric predictors and ROCIO+\_CAN target variables used in this study.}
\label{tab:Variables}
\end{center}
\end{table}

The dataset covers the temporal period 1990--2022. For model development, the years 2011--2012 are reserved as the validation set, while the remaining data are divided into a training set (1990--2010) and test set (2013--2022), as described in the training strategy presented in the \nameref{sec:TrainingStrat} subsection.

\subsection{Latent generative downscaling}

In this study, two distinct generative modeling paradigms operating in a compressed latent space are evaluated: denoising diffusion probabilistic models (DDPMs) \citep{ho2020denoisingdiffusionprobabilisticmodels} and rectified flow (RF) \citep{liu2022flow}.  To ensure consistency and allow for a direct comparison, both generative frameworks are built on two shared core components: a variational autoencoder (VAE) which performs the spatial compression, and a Transformer-based neural backbone (ClimDiT) that drives the generative process. This section describes the latent space formulation, the mathematical frameworks governing the generative processes, and the shared network architecture.

\subsubsection{The latent space: variational autoencoder (VAE)}\label{sec:TheVAE}

Both generative frameworks use the same underlying VAE to map the high-resolution multivariate target fields into a compact latent representation before generative modeling. This design reduces the spatial dimensionality processed by the denoising backbone, preserves cross-variable dependencies within a shared representation, and reduces the computational cost of generating large ensembles. As shown by the ablation experiments in Appendix A, this regularized latent representation also helps to reduce the number of denoising steps, thereby accelerating the generative process and lowering the sampling cost.

The VAE follows the architecture used in the SDXL latent-diffusion framework \citep{podell2024sdxl}, which builds on the latent autoencoding strategy of \citet{rombach2022ldm}. The approach adopted here is the same as that described in \cite{EsquivelGonzalez2026climdit}. Prior to encoding, precipitation is transformed using \(\log1p(x)\) to reduce skewness and to make its distribution more compatible with the smooth, image-like fields for which this family of autoencoders was originally trained. Additionally, the three variables are normalized to the [-1,1] interval.  The encoder then compresses the three high-resolution target channels (daily maximum temperature, daily minimum temperature, and log-transformed precipitation) into a four-channel latent tensor with a spatial downsampling factor of 8. This reduces the spatial dimensions from $72\times192$ grid points to $9\times24$ latent grid points. Ultimately, the decoder then reconstructs the physical target fields from the generated latent representations. During autoencoder training, the latent representations are produced by the encoder. However, for the downscaling task, the generative model, either Gaussian diffusion (GD) or rectified flow (RF), is trained to generate these latent representations conditioned on the low-resolution synoptic-scale predictors.


To adapt the latent representation to the specific climatological characteristics of the Canary Islands, the VAE is fine-tuned using a composite loss function that balances spatial reconstruction fidelity with a Kullback-Leibler (KL) regularization term ($\mathcal{L}_{\text{reg}}$):
\begin{equation}
    \mathcal{L}_{\text{AE}} = \text{MSE}(\mathcal{D}(\mathcal{E}(x)), x) + \lambda \cdot \mathcal{L}_{\text{reg}}(x, \mathcal{E})
\end{equation}
where $\mathcal{E}$ represents the encoder, $\mathcal{D}$ denotes the decoder, and $\text{MSE}$ is the mean squared error responsible for minimizing reconstruction discrepancies. The $\mathcal{L}_{\text{reg}}$ term penalizes deviations of the approximate latent distribution from a standard normal distribution $\mathcal{N}(0, \mathbf{I})$. This regularization is crucial, as it ensures a continuous and smooth latent space. The coefficient $\lambda$ controls the contribution of this regularization term and is set to $10^{-8}$, following previous studies \citep{EsquivelGonzalez2026climdit}.

Rather than training the autoencoder from scratch for the Canary Islands domain, the network is initialized using weights obtained from a previous downscaling study over the Iberian Peninsula \citep{EsquivelGonzalez2026climdit}. Those weights correspond to an SDXL-VAE adapted and pretrained on the Iberian Peninsula downscaling task using high-resolution ROCIO-IBEB targets. This transfer-learning step provides a climate-aware latent space that has already learned physically meaningful spatial structures for the same three target variables.

\begin{figure}[htbp]
    \centering
    \includegraphics[width=0.9\textwidth]{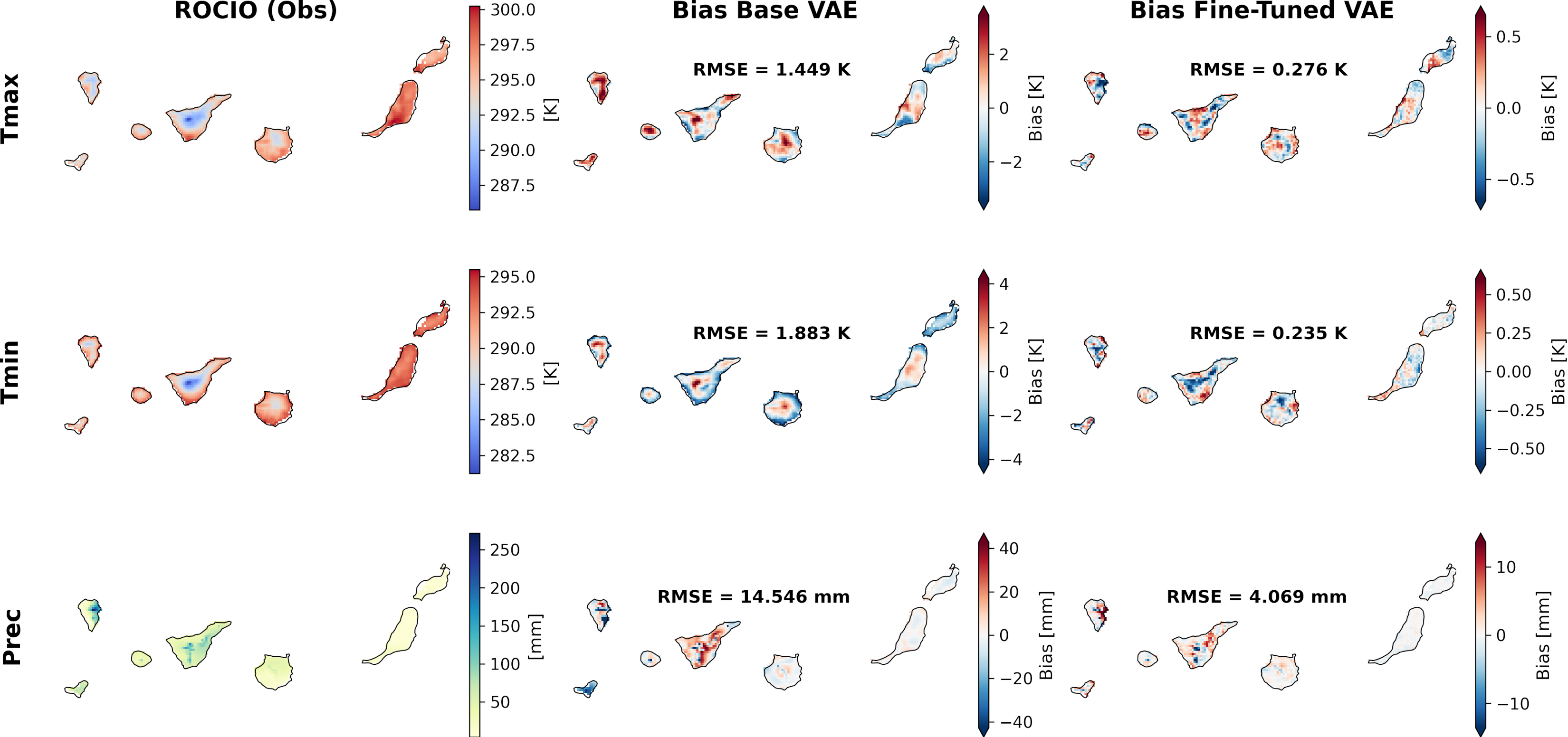}
    \caption{Comparison of the high-resolution ground-truth fields and the corresponding variational autoencoder (VAE) reconstructions for a randomly selected day with intense precipitation.}
    \label{fig:vae_recons}
\end{figure}

Figure \ref{fig:vae_recons} illustrates the quality of this latent representation by comparing a randomly selected rainy sample of the high-resolution fields with the corresponding VAE reconstructions, both before (Base VAE) and after fine-tuning (Fine-Tuned VAE). The reconstructed temperature and precipitation field errors are reduced by approximately one order of magnitude after fine-tuning, achieving the level of accuracy required to train a generative model in this latent space.

Additionally, Table \ref{tab:TabResults_AE} reports the mean reconstruction errors on the test set along with the latent space statistics ($Z_{\text{mean}}$ and $Z_{\text{std}}$) for the different VAE weight configurations, confirming the trend observed in Figure \ref{fig:vae_recons}. Interestingly, the original SDXL weights achieve lower reconstruction errors for both temperature variables than the Base VAE, although they also produce a less regularized latent space. This trade-off makes the choice of training initialization non-trivial. The Base VAE, pretrained on the ROCIO-IBEB dataset, became highly specialized in reconstructing the spatial masking patterns of the Iberian Peninsula, which deteriorated its performance in a new masked domain. Conversely, the original SDXL weights prioritize unconstrained reconstruction over latent space normality, yielding good results in temperature fields. However, since the Base VAE is already adapted to geophysical fields, including features such as precipitation intermittency, and provides stronger latent regularization, it is selected as the starting point for subsequent fine-tuning instead of the generic SDXL image-domain representation.

\begin{table}
\begin{center}
\begin{tabular}{lccccc}
\topline
& \multicolumn{3}{c}{\textbf{Reconstruction error (RMSE)}}
& \multicolumn{2}{c}{\textbf{Latent-space statistics}} \\
\midline
\textbf{Weight configuration}
& \textbf{Tmax}
& \textbf{Tmin}
& \textbf{Prec}
& $\boldsymbol{Z_{\mathrm{mean}}}$
& $\boldsymbol{Z_{\mathrm{std}}}$ \\
\midline
Fine-tuned VAE
& 0.15
& 0.13
& 0.29
& $6.5 \times 10^{-2}$
& 0.98 \\

Base VAE (ROCIO-IBEB)
& 2.05
& 2.50
& 1.51
& $1.4 \times 10^{-3}$
& 0.84 \\

Original weights (SDXL)
& 1.27
& 1.19
& 2.41
& $6.6 \times 10^{-2}$
& 9.96 \\
\botline
\end{tabular}
\caption{Test-set autoencoder results for different model weight configurations.}
\label{tab:TabResults_AE}
\end{center}
\end{table}

\subsubsection{Generative paradigms: Gaussian diffusion and rectified flow}

Once the high-resolution targets are compressed, the generative task consists of learning to map a simple noise distribution to the complex latent data distribution, conditioned on the low-resolution ERA5 synoptic scale predictors. To this end, two distinct mathematical frameworks are evaluated: Gaussian diffusion and rectified flow. To remain consistent with the standard notation used in the literature for each paradigm, the notation differs slightly between the two formulations: Gaussian diffusion denotes the clean data by $x_0$ and the pure noise by $x_T$, whereas rectified flow denotes the noise by $X_0$ and the clean data by $X_1$.

\paragraph{Gaussian diffusion (DDPM)}
The first paradigm follows the denoising diffusion probabilistic models (DDPM) formulation. Diffusion is modeled as a discrete-time Markov chain that progressively corrupts the data, converting the original data distribution $q(x_0)$ into a Gaussian distribution $q(x_T) \sim \mathcal{N}(0, \mathbf{I})$. Given an initial clean sample $x_0$, the forward process is defined by $q(x_t | x_{t-1}) = \mathcal{N}(x_t ; \sqrt{1 - \beta_t} x_{t-1} , \ \beta_t \mathbf{I})$ for $t = 1, 2, ..., T$, with the closed form:
\begin{align}
q(x_t | x_0) &= \mathcal{N}(x_t ; \sqrt{\bar{\alpha}_t} x_0 , (1 - \bar{\alpha}_t) \mathbf{I}) \\
x_t &= \sqrt{\bar{\alpha}_t }x_0 +\sqrt{1 - \bar{\alpha}_t}\epsilon ,\quad \epsilon \sim \mathcal{N}(0, \mathbf{I})
\end{align}
where $\alpha_t = 1 - \beta_t $ and $\bar{\alpha}_t = \prod_{s=1}^t\alpha_s$ are controlled by a predefined linear variance schedule $\{\beta_t\}$.

The objective is to approximate the reverse Markov chain, conditioned on the low-resolution synoptic maps (cond):
\begin{equation}
	p_\theta(x_0 | \text{cond}) = p_\theta(x_T) \prod_{t = 1}^T p_\theta(x_{t-1} | x_{t}, \text{cond})
\end{equation}

The neural network learns to predict the noise term $\epsilon_\theta$ and the variance $\Sigma_\theta$ required to estimate the true posterior distribution. Following previous work \citep{ADM_NEURIPS2021_49ad23d1, peebles2023dit}, the network is trained to optimize the variational lower bound by minimizing the simple noise reconstruction loss ($\mathcal{L}_{\text{simple}}$) for $\epsilon_\theta$ and the Kullback-Leibler divergence ($\mathcal{L}(\theta)$) for $\Sigma_\theta$.

\paragraph{Rectified flow (RF)}
As an alternative to stochastic Markov chains, rectified flow (RF) is implemented. RF is an ordinary differential equation (ODE)-based generative framework that learns a transport map between a standard Gaussian noise distribution ($X_0 \sim \mathcal{N}(0, \mathbf{I})$) and the target data distribution ($X_1$). RF defines the following linear interpolation path:
\begin{equation}
    X_t = t X_1 + (1 - t) X_0, \quad t \in [0, 1]
\end{equation}

Instead of predicting the noise, the neural network $v_\theta(X_t, t, \text{cond})$ is trained to estimate the constant velocity vector along this path, $v = X_1 - X_0$, conditioned on the atmospheric predictors. During training, time steps $t$ are sampled from a Logit-Normal distribution to improve the learning dynamics \citep{hu2025improvingrectifiedflowboundary}.

To enforce consistency with the target and noise spaces at the endpoints of the generative process, particularly when sampling different initial noise realizations at inference time, a boundary-enforced velocity mechanism is applied \citep{hu2025improvingrectifiedflowboundary}. The raw network prediction $v_{\text{raw}}$ is modulated using a trigonometric mask so that the boundary conditions $v(X_0, 0) = -X_0$ and $v(X_1, 1) = X_1$ are satisfied:
\begin{equation}
    v(X_t, t) = \cos\left(\frac{\pi}{2} t\right) (-X_t) + \sin\left(\frac{\pi}{2} t\right) (X_t) + \sin(\pi t) v_{\text{raw}}
\end{equation}
During inference, realizations are generated by integrating the resulting ODE using a numerical solver, such as the explicit Euler method. 

\subsubsection{Backbone architecture: ClimDiT}\label{sec:Arch}

Regardless of the generative paradigm employed, Gaussian diffusion or rectified flow, the core learning task is conceptually similar: learning to predict a time-dependent vector field, either the added noise or the transport velocity, conditioned on external atmospheric predictors. To accomplish this, the Climate Diffusion Transformer (ClimDiT) architecture, described in \cite{EsquivelGonzalez2026climdit}, is used as the shared neural backbone. 

Building on the original Diffusion Transformer (DiT) framework, ClimDiT uses self-attention mechanisms to model complex long-range spatial dependencies. The architecture consists of a main generative network and a dedicated conditioning encoder.

To ensure that high-resolution generation remains physically consistent with daily synoptic weather patterns, the model processes the coarse-resolution ERA5 predictors through a conditioning encoder. This encoder consists of a dedicated Vision Transformer-style feature extractor with 6 transformer blocks and approximately 16 million parameters, which infers long-range spatial relationships among the synoptic patch tokens across the entire domain. The resulting features are combined with the current generative timestep through adaptive layer normalization (adaLN-Zero), dynamically adapting the context to the noise level or mass transport interpolation state throughout the reverse generative process.  Finally, the spatially aware synoptic features are finally aggregated into a unified conditioning vector using a mixed pooling strategy that combines the mean and maximum values across all 'conditioning' tokens. This vector is then used to modulate the main generative blocks, which process noise using again adaLN-Zero.  

Additionally, the network's capacity is optimized in this study. The total number of model parameters is reduced from 72 million to 43 million by decreasing the number of attention heads from eight to six and adjusting the associated architectural dimensions. To avoid an excessive reduction in the conditioning pathway, which must extract a synoptically rich representation from the ERA5 predictors, the conditioning encoder is instead slightly enlarged from 5 to 6 transformer blocks. At the same time, the patch size used to tokenize the compressed latent space is reduced from  $3 \times 3$ to a $2 \times 2$. This finer tokenization facilitates the accurate capture of spatial relationships that cannot be accounted for when using a larger patch size. This helps the network to better model the highly localised microclimates characteristic of regions such as the Canary Islands, which are also small in spatial terms ($9\times24$ in latent space). Some preliminary experiments, conducted prior to the introduction of training monitoring using a validation set, such as those in \citet{EsquivelGonzalez2026climdit}, did not show performance degradation using this configuration, despite the reduction in model complexity.

\subsubsection{Sampling and ensemble configurations}

For the Gaussian diffusion model, individual samples are generated by starting from Gaussian noise and recursively applying the learned reverse transitions using a linear variance schedule ranging from $10^{-4}$ to $10^{-2}$. Although the previous ClimDiT application over the Iberian Peninsula used 1000 denoising steps \citep{EsquivelGonzalez2026climdit}, internal validation experiments showed that the regularized latent space allows ClimDiT to produce stable and comparable results using only 100 sampling steps, selected at evenly spaced intervals from the 1,000-step diffusion process used during training. To ensure this reduction was not an artifact of the new spatial domain, this 100-step setting was re-tested alongside the new architectural and training updates (see subsections \nameref{sec:Arch} and \nameref{sec:TrainingStrat}) on the original Iberian Peninsula setup, achieving results equal to or slightly better than the previous ClimDiT configuration. In addition to individual stochastic realizations, an ensemble configuration is evaluated by running the trained ClimDiT model 50 times for each input condition, using a different initial noise realization in each run. This ensemble is denoted ClimDiT-ENS.

For Rectified flow, samples are generated by integrating the learned transport field. ClimDiT-RF model is also evaluated using 100 integration steps, therefore using the same sampling budget than ClimDiT. Both deterministic Euler integration and a stochastic curved Euler sampler are tested. The latter injects controlled noise during numerical integration to mitigate truncation errors and increase sample diversity. Because the stochastic sampler produces better overall scores, it is adopted for the RF ensemble experiments. The corresponding ensemble configuration, denoted as ClimDiT-RF-ENS, is obtained by running the RF model 50 times for each input condition, with a different initial noise realization in each run. Increasing the number of integration steps from 100 to 1000 produces a small additional improvement, although the gain is insufficient to alter the conclusions of the main comparison. Conversely, further reductions in the number of RF integration steps could, in principle, be achieved with iterative trajectory-straightening strategies such as ReFlow \citep{liu2022flow}. However, these additional distillation stages are not explored here because the 100-step RF configuration does not achieve the expected performance. Reducing the number of integration steps is therefore unlikely to address the main source of error in this setting.

\subsection{Benchmark models}
To contextualize the performance of the generative models, a comprehensive set of deterministic and probabilistic convolutional neural network (CNN) baselines is included. These benchmark models use loss functions tailored to the specific climate variable being downscaled, following the optimal combinations of architectures and objective functions identified in previous reviews \cite{AreDeepLearningMeth}.

\subsubsection{Univariate baselines}
For both maximum and minimum temperature, the best-performing configurations use the DeepESD architecture trained with either a mean squared error loss (DeepESD-MSE) or a negative log-likelihood (NLL) loss (DeepESD-STO). The deterministic DeepESD-MSE directly minimizes the average squared difference between predictions and observations, whereas the probabilistic DeepESD-STO minimizes the NLL by predicting the mean and standard deviation of a Gaussian distribution at each grid point.

For precipitation, accounting for the highly skewed distribution and the large proportion of dry days requires specialized approaches. Therefore, DeepESD trained with an asymmetric loss function (DeepESD-ASYM) and a U-Net architecture with a stochastic output formulation (UNet-STO) are selected as benchmark models. 

The asymmetric loss function is designed to penalize underestimations of extreme values and is defined as: 
\begin{equation}
    \mathcal{L} = \frac{1}{N}\sum_{i=1}^N |y_i - \hat{y}_i| + \gamma^2 \cdot \max(0, y_i - \hat{y}_i) 
\end{equation}
where $\gamma = G(y)$, with $G$ corresponding to the cumulative distribution of a gamma function fitted for each grid point in the historical dataset.

Conversely, the stochastic loss for precipitation (UNet-STO) consists of minimizing the NLL of a Bernoulli-Gamma distribution:
\begin{equation}\label{eq:BernGammaDist}
    P(y|X;p,\alpha,\beta) = \begin{cases}
        1 - p & \text{if } y < 0.1 \\
        \frac{p}{\Gamma (\alpha)\beta^\alpha}(y-0.1)^{\alpha-1}e^{\frac{-(y-0.1)}{\beta}} & \text{if } y \ge 0.1 
    \end{cases} 
\end{equation}
where $p$ is the probability of rain, $\alpha$ and $\beta$ are the shape and scale parameters of the gamma distribution, and $\Gamma$ is the Gamma function.

\subsubsection{Multivariate baseline}
While the benchmarks described above operate on a univariate basis, the proposed generative frameworks (ClimDiT and ClimDiT-RF) are inherently multivariate. To broaden the comparison to include other multivariate baselines, multivariate versions of both DeepESD (DeepESD-MULT) and U-Net (UNet-MULT) are also evaluated. These models adapt the corresponding CNN architectures to produce all three target variables simultaneously: maximum temperature, minimum temperature, and precipitation. To ensure consistency, both multivariate baselines are trained using the same target-variable preprocessing and validation strategy as the generative models (see Subsections \nameref{sec:TheVAE} and \nameref{sec:TrainingStrat}). Since UNet-MULT consistently outperformed DeepESD-MULT across the test diagnostics, only UNet-MULT is included in the main results, while the DeepESD-MULT results are reported in Appendix A.

Table \ref{tab:Summary_models} summarizes the set of models evaluated in this study. For each target variable, the inclusion of both deterministic and stochastic baselines, together with univariate and multivariate architectures, provides a robust framework for benchmarking the proposed diffusion and rectified flow models.

\begin{table}
\begin{center}
\begin{tabular}{lccc}
\topline
\textbf{Model}
& \textbf{Predicted variables}
& \textbf{Loss / objective}
& \textbf{Parameters (M)} \\
\midline
DeepESD-MSE
& \(T_{\max}\) or \(T_{\min}\)
& Mean squared error
& 11 \\

DeepESD-STO
& \(T_{\max}\) or \(T_{\min}\)
& Gaussian NLL
& 22 \\

DeepESD-ASYM
& Precipitation
& Asymmetric loss
& 1 \\

UNet-STO
& Precipitation
& Bernoulli--Gamma NLL
& 8 \\

UNet-MULT
& \(T_{\max}\), \(T_{\min}\), precipitation
& Mean squared error
& 8 \\

DeepESD-MULT
& \(T_{\max}\), \(T_{\min}\), precipitation
& Mean squared error
& 33 \\
\midline
ClimDiT
& \(T_{\max}\), \(T_{\min}\), precipitation
& Diffusion loss
& 43 + 84 VAE \\

ClimDiT-ENS
& \(T_{\max}\), \(T_{\min}\), precipitation
& Diffusion loss
& 43 + 84 VAE \\

ClimDiT-RF
& \(T_{\max}\), \(T_{\min}\), precipitation
& ODE velocity matching
& 43 + 84 VAE \\

ClimDiT-RF-ENS
& \(T_{\max}\), \(T_{\min}\), precipitation
& ODE velocity matching
& 43 + 84 VAE \\
\botline
\end{tabular}
\caption{Summary of the statistical downscaling models evaluated in this study, including their predicted variables, objective functions, and number of parameters.}
\label{tab:Summary_models}
\end{center}
\end{table}

\subsection{Training strategy and validation metric}\label{sec:TrainingStrat}
Internal experiments showed that the  models were prone to developing larger spatial errors in extreme values during prolonged training over the Canary Islands domain. To mitigate this issue, an early-stopping and model-selection strategy is implemented using a custom validation metric evaluated on the 2011--2012 validation set. This composite metric is computed independently for each normalized target variable (Tmax,Tmin and Log1p(Prec)) and then averaged across variables. For variable $v$, the metric combines the spatially averaged temporal root mean squared error (RMSE) with the spatial RMSE of the 2nd ($p02$) and 98th ($p98$) percentiles:
\begin{align}
    \text{Val}_{\text{score}}
    &= \frac{1}{V}\sum_{v=1}^{V}
    \left[ \overline{\text{RMSE}_v} + w\left(\text{PE}_{p98,v} + \text{PE}_{p02,v}\right)\right], \\
    \overline{\text{RMSE}_v}
    &= \frac{1}{G}\sum_{i=1}^{G}
    \sqrt{\frac{1}{T}\sum_{t=1}^{T}\left(\hat{y}_{t,v,i}-y_{t,v,i}\right)^2}, \\
    \text{PE}_{pq,v}
    &= \sqrt{\frac{1}{G}\sum_{i=1}^{G}\left(\widehat{pq}_{v,i}-pq_{v,i}\right)^2},
    \qquad pq \in \{p02,p98\}.
\end{align}
where $V=3$ denotes the number of variables, $G$ the number of grid cells, $T$ the number of validation days, $\widehat{p02}_{v,i}$ and $\widehat{p98}_{v,i}$ are the predicted temporal percentiles at each grid cell, and $w$ is a weighting hyperparameter. The first term, $\overline{\text{RMSE}_v}$, applies an arithmetic mean over space, giving equal weight to the temporal errors at all grid cells. By contrast, the second term, $\text{PE}_{pq,v}$, is computed as a spatial root mean square, so that the grid cells with the largest errors at the extreme percentiles are penalized more heavily.

\begin{figure}[h]
    \centering
    \includegraphics[width=0.8\textwidth]{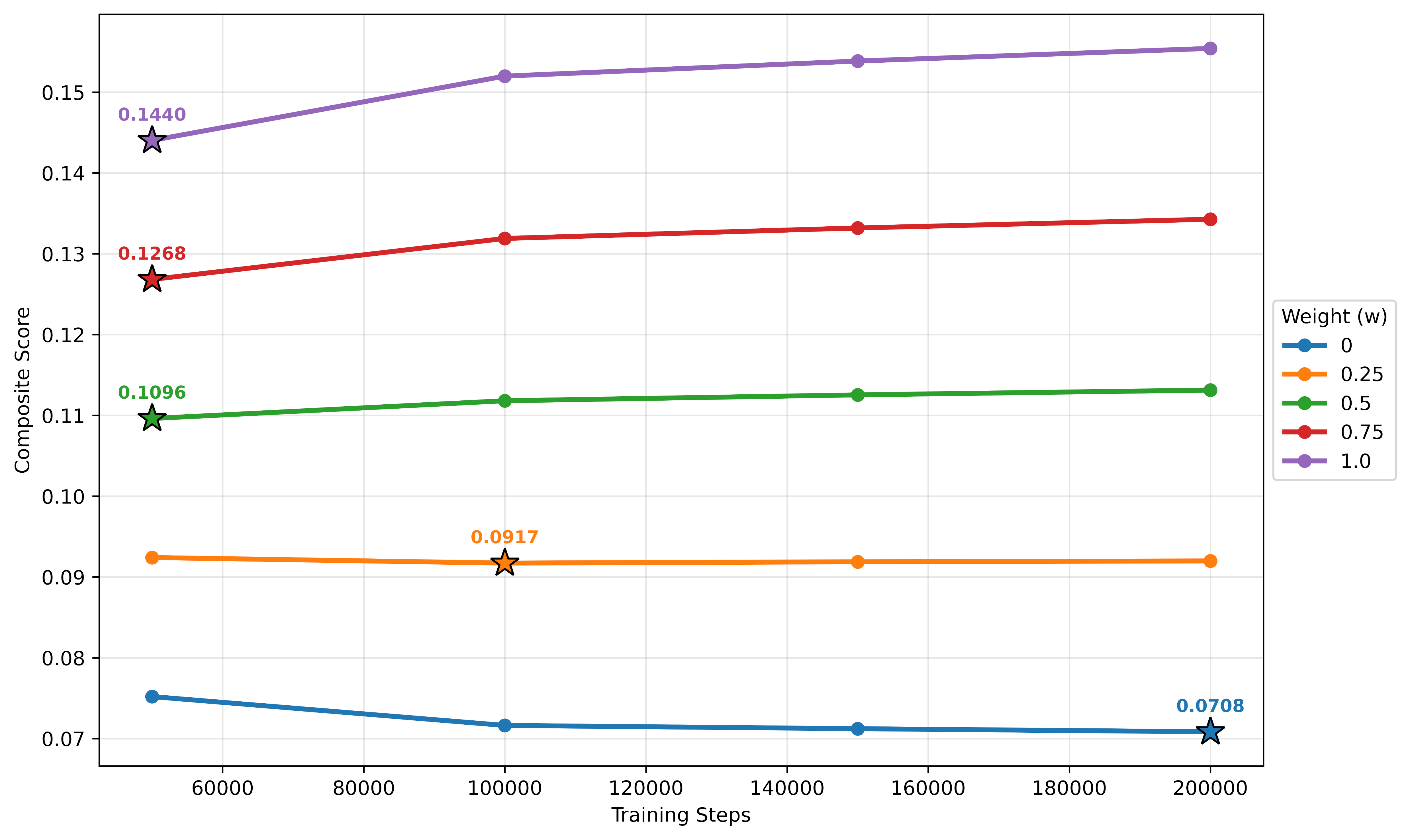}
    \caption{Validation score as a function of training step. Each curve corresponds to a different weight ($w$) assigned to the extreme-percentile term in the composite validation metric. The minimum value of each curve, which would correspond to the selected model checkpoint, is also marked.}
    \label{fig:WeightsSensitivity}
\end{figure}

To avoid over-penalizing the model for stochastic fluctuations in extreme values, the weight assigned to the extreme-error term was empirically assessed. As shown in Figure \ref{fig:WeightsSensitivity}, when this term is excluded ($w = 0$), the model considered here, ClimDiT, continues to improve in terms of RMSE throughout training, while the errors in the extreme percentiles follow the opposite trend. This divergence provides the main motivation for the composite criterion: the checkpoint selected on the basis of average error alone is not necessarily the one that best preserves the distributional tails. For $w \ge 0.5$, the composite validation metric reaches its minimum at the first validation step, indicating that subsequent improvements in RMSE are accompanied by a deterioration in the representation of extremes. Because the validation set spans only two years and extreme-errors estimates are noisier than RMSE estimates, $w = 0.5$ was selected as the best compromise between overall accuracy and the representation of extreme values.

The same composite validation metric was also evaluated for the multivariate CNN baselines. For both DeepESD-MULT and UNet-MULT, checkpoint selection based on this metric led to better overall performance than selection based on RMSE alone. Therefore, the multivariate benchmark configurations reported in the main text use this composite model-selection criterion. A direct comparison with the corresponding variants selected without the composite metric is provided in Appendix A.

Regarding the training hyperparameters, the benchmark models were trained using the Adam optimizer with a learning rate of $10^{-4}$ and a batch size of 64. Early stopping was applied after 60 epochs without improvement in the composite validation metric. The VAE and generative models were trained using AdamW with the same batch size, 64, and learning rate, $10^{-4}$, and with zero weight decay. The VAE was trained for 350 epochs, retaining the checkpoint with the lowest validation loss. The generative models were trained for 200,000 steps, corresponding to approximately 1,650 epochs, using an exponential moving average (EMA) of the model weights with a decay rate of 0.9999. Their validation performance was evaluated every 50,000 steps ($\sim 400$ epochs).


\section{Results}\label{sec:Results}

This section presents the evaluation of the proposed generative downscaling models and the benchmark baselines over the test period. The analysis first examines spatial coherence through SSIM and RAPSD diagnostics, then evaluates deterministic, probabilistic, and multivariate performance for temperature and precipitation, assesses the role of the latent space, and finally compares the computational efficiency and memory requirements of the different approaches. For clarity, the definitions and mathematical formulations of the main metrics used throughout this section are provided in the glossary in Appendix B. Ensemble models use the average of the 50 members for deterministic metrics and the empirical cumulative distribution for probabilistic ones. 

\subsection{Spatial coherence}
To assess the structural realism and spatial fidelity of the downscaled fields, the structural similarity index (SSIM) and the radially averaged power spectral density (RAPSD) are evaluated. SSIM quantifies the structural similarity between the predicted and observed high-resolution fields, whereas RAPSD characterizes the spatial distribution of energy and variance across different spatial scales (wavelengths).

\begin{table}
\begin{center}
\begin{tabular}{lccc}
\topline
\textbf{Model}
& \textbf{\(T_{\min}\)}
& \textbf{\(T_{\max}\)}
& \textbf{\(\log(1+\mathrm{Prec})\)} \\
\midline
ClimDiT
& 0.961
& 0.954
& 0.782 \\

ClimDiT-ENS
& 0.965
& 0.960
& 0.788 \\

ClimDiT-RF
& 0.963
& 0.956
& 0.791 \\

ClimDiT-RF-ENS
& 0.965
& 0.959
& 0.794 \\
\midline
UNet-MULT
& 0.955
& 0.941
& 0.713 \\

DeepESD-MSE
& 0.965
& 0.955
& -- \\

DeepESD-STO
& 0.807
& 0.790
& -- \\

DeepESD-ASYM
& --
& --
& 0.728 \\

UNet-STO
& --
& --
& 0.603 \\
\botline
\end{tabular}
\caption{Mean SSIM values over the test set for each model and target variable. The optimal SSIM value is 1.}
\label{tab:SSIM}
\end{center}
\end{table}

Table \ref{tab:SSIM} shows the daily averaged SSIM for each model and target variable. For minimum ($T_{min}$) and maximum ($T_{max}$) temperature, the deterministic baselines (DeepESD-MSE and UNet-MULT) and the proposed generative models (ClimDiT and ClimDiT-RF, together with their respective ensembles) achieve near-optimal SSIM values, indicating an accurate reconstruction of the observed spatial patterns. By contrast, the stochastic baseline (DeepESD-STO) exhibits noticeably lower SSIM values for both temperature variables, suggesting reduced spatial coherence, likely due to grid-point-wise stochastic sampling.

In the case of precipitation, evaluated on $log1p$-transformed fields to account for the highly skewed distribution of the variable, the performance differences among the methods become more pronounced. Both univariate baselines struggle to reproduce the spatial structure accurately, with UNet-STO presenting the lowest structural similarity. The generative models (ClimDiT and ClimDiT-RF), and their ensemble variants, achieve the highest SSIM scores, outperforming both the univariate baselines and the deterministic multivariate U-Net. These results indicate that generative frameworks are better able to preserve complex, fine-scale precipitation structures.

\begin{figure}
    \centering
    \includegraphics[width=\linewidth]{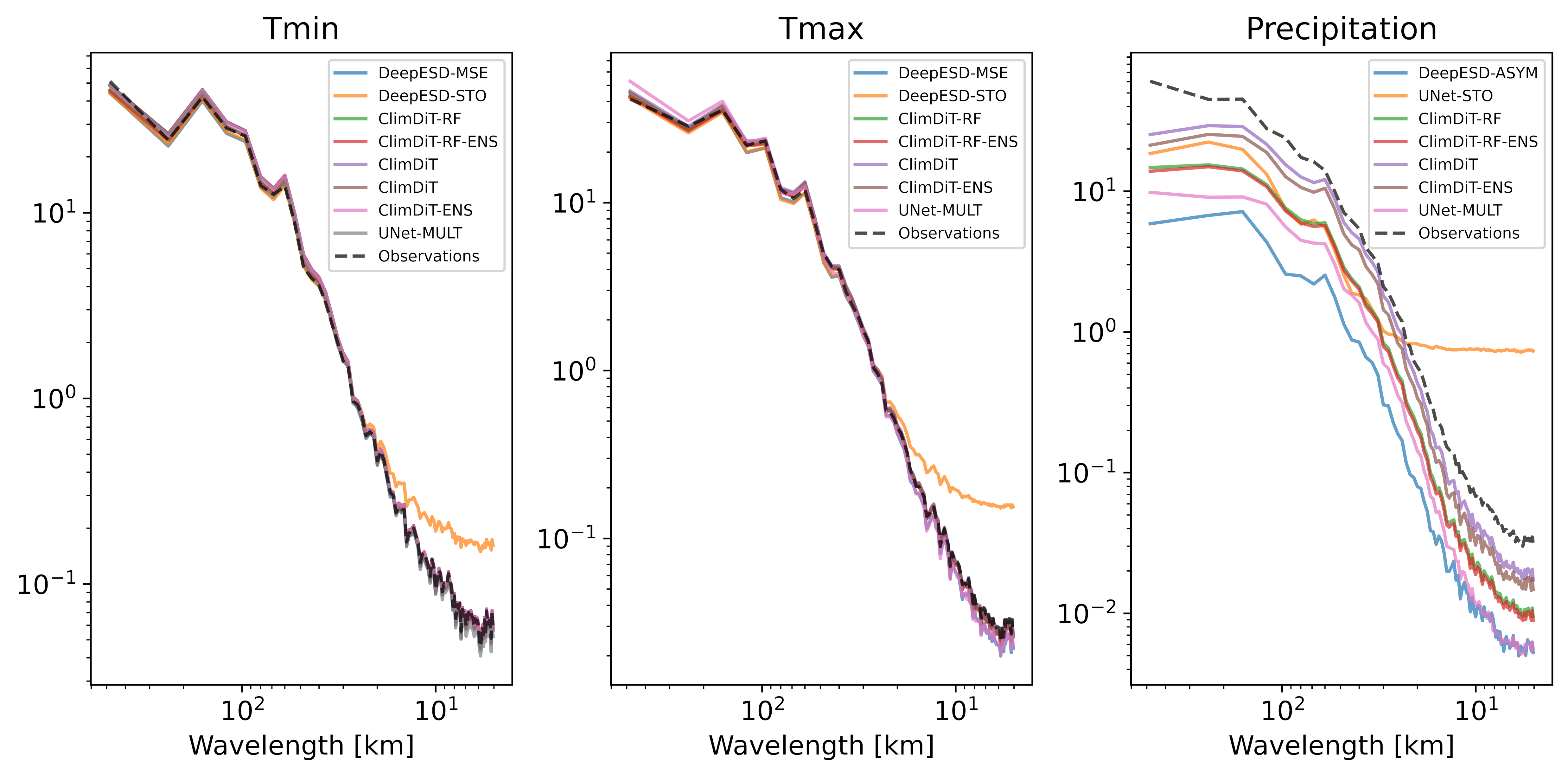}
    \caption{Radially averaged power spectral density (RAPSD) curves. Models and observations are shown.}
    \label{fig:RAPSD}
\end{figure}

Figure \ref{fig:RAPSD}  provides a scale-by-scale analysis of this spatial realism. For $T_{min}$ and $T_{max}$, the spectral curves of most models closely with follow observational spectrum at synoptic and meso-alpha scales, corresponding to wavelengths larger than $>100$ km. However, at finer scales, below 50 km, the stochastic baseline (DeepESD-STO) exhibits excess spectral power. This deviation suggest the presence of artificial high-frequency variability associated with the unconstrained pixel-by-pixel Gaussian sampling. By contrast, the generative models and deterministic baselines closely reproduce the observed spectrum across the full range of wavelengths.

For precipitation, the spectral differences are particularly pronounced. The deterministic models (DeepESD-ASYM and UNet-MULT) substantially underestimate the observed spectrum across all spatial scales. This behaviour is consistent with the well-known tendency of regression-based methods to produce overly smooth fields when minimizing grid-point-wise errors. At the finest scales, the stochastic UNet-STO model exhibits a flattened spectral tail, suggesting the presence of artificial grid-scale noise rather than physically coherent small-scale variability. Notably, the rectified flow models (ClimDiT-RF and ClimDiT-RF-ENS) also show a pronounced underestimation of spectral power, although their performance remains better than that of the deterministic baselines.

By contrast, the Gaussian diffusion models (ClimDiT and ClimDiT-ENS) are the most effective at mitigating the loss of spatial variance. These models recover a substantial fraction of the observed spectral power and follow the observational spectrum more closely than the benchmark models across the full range of scales, from large synoptic wavelengths to the finer mesoscale range. These results suggest that, although both generative frameworks operate within a compressed latent space, the progressive denoising process used by the diffusion-based approach is particularly effective at reproducing realistic fine-scale precipitation variability while preserving large-scale spatial coherence. This behaviour reduces the spatial decoupling and over-smoothing observed in the other approaches. Appendix A presents an ablation study showing that the Gaussian diffusion formulation, rather than the VAE alone, is the main contributor to these improvements.

\subsection{Deterministic and multivariate performance}
In this subsection, the spatial distributions of the deterministic and probabilistic error metrics, computed at each high-resolution land grid cell across the Canary Islands domain, are analyzed using violin plots. This visualization highlights not only the median performance of each model but also its robustness and consistency across the complex topography of the archipelago.

\subsubsection{Minimum temperature downscaling}

\begin{figure}
    \centering
    \includegraphics[width=\linewidth]{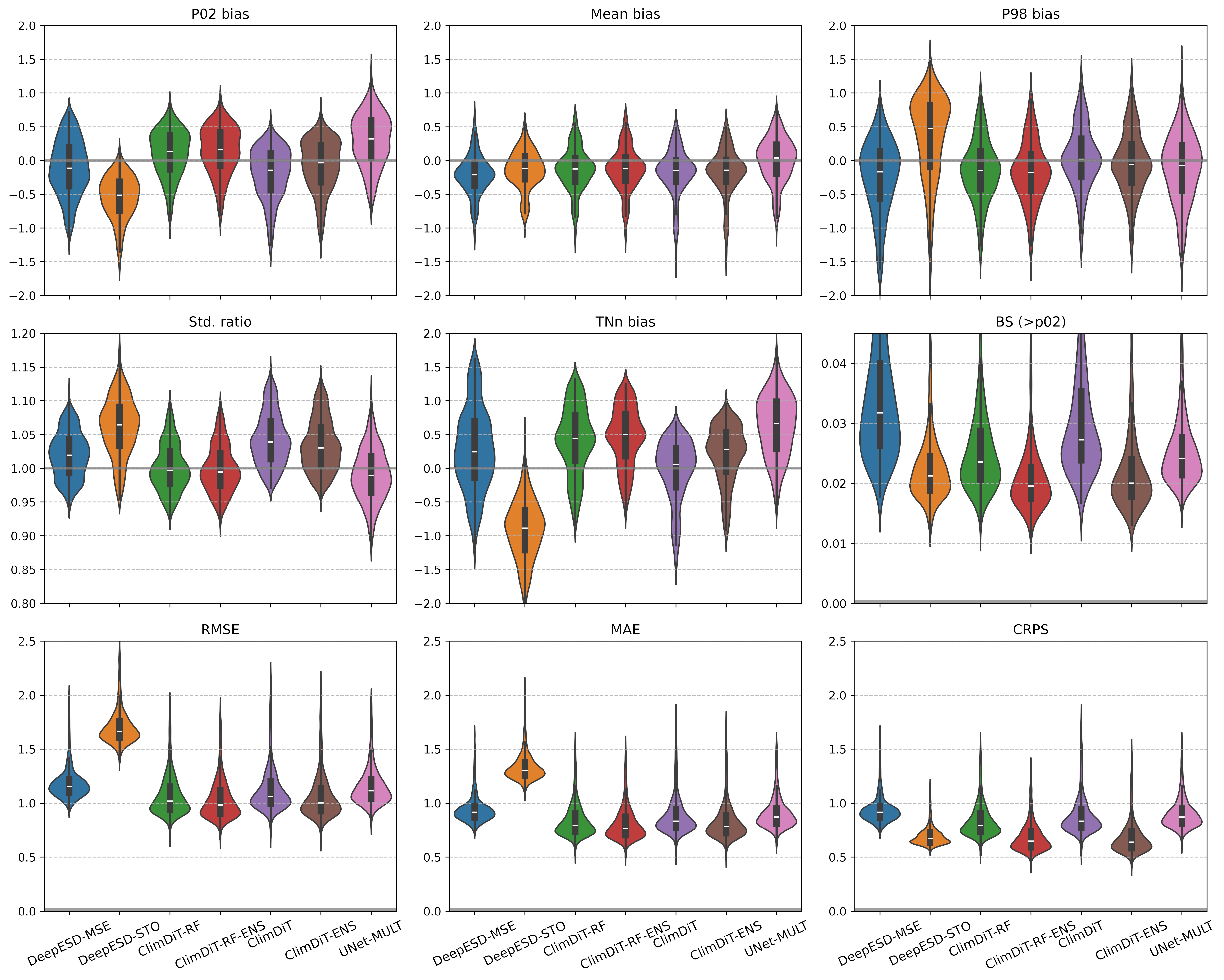}
    \caption{Violin plots of metrics for minimum temperature}
    \label{fig:ViolinMinTemp}
\end{figure}

Figure \ref{fig:ViolinMinTemp} presents the distribution of evaluation metrics for daily minimum temperature ($T_{min}$). The point-wise deterministic errors, specifically the root mean square error (RMSE) and mean absolute error (MAE), show that both the generative frameworks (ClimDiT and ClimDiT-RF) and the deterministic UNet-MULT achieve the lowest (best) values. These are closely followed by the DeepESD-MSE baseline. Conversely, the stochastic baseline (DeepESD-STO) exhibits a substantially higher deterministic errors than the other models, reflecting the expected trade-off of its unconstrained probability-matching objective and stochastic sampling strategy.

\begin{figure}
    \centering
    \includegraphics[width=0.9\linewidth]{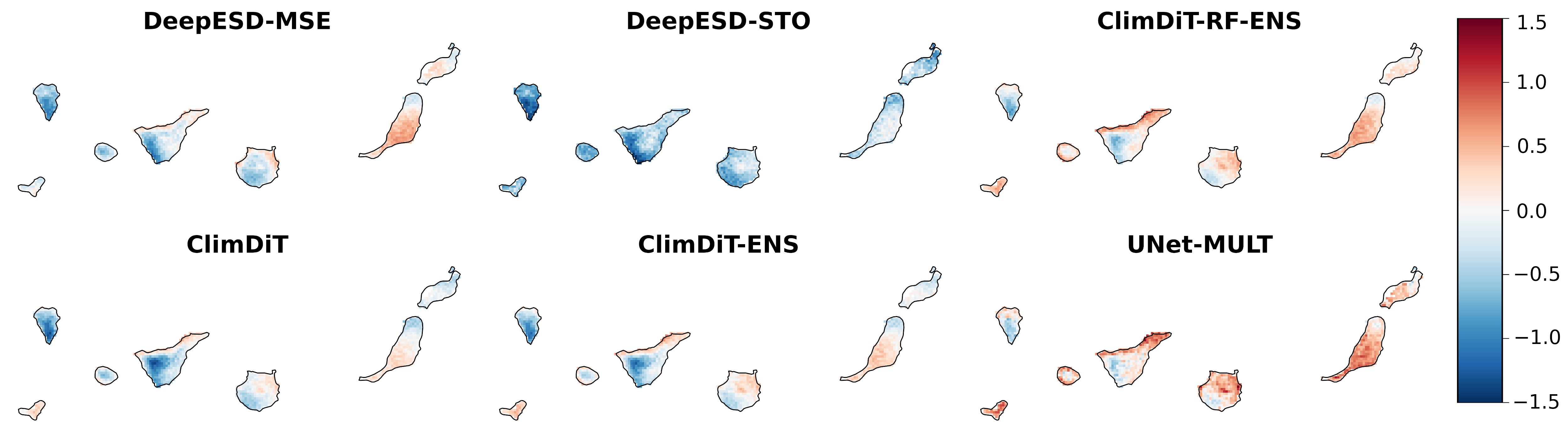}
    \caption{Maps of the bias in the 2nd percentile (p02) of daily minimum temperature (K) over the test period. ClimDiT-RF is not included but its 50-member ensemble, ClimDiT-RF-ENS, is shown instead.}
    \label{fig:p02_tmin_map}
\end{figure}
The analysis of climatological biases reveals distinct model behavior across the distribution tails. For the mean bias, performance is generally comparable across models; UNet-MULT is well-centered around zero but exhibits a larger spatial dispersion, as indicated by its wider interquartile range and more elongated violin shape. Regarding the cold extremes ($p02$ bias), all models remain relatively close to the ideal value of zero. However, UNet-MULT slightly overestimates this cold tail, whereas DeepESD-STO shows a larger bias characterized by an underestimation centered around $-0.5 K$. The spatial maps in Figure \ref{fig:p02_tmin_map} further show that this underestimation by DeepESD-STO extends across much of the archipelago, while the overestimation by UNet-MULT is also spatially widespread. The generative models and DeepESD-MSE exhibit smaller and more spatially localized biases, although some underestimation remains over parts of the  western islands. Although the spatial distribution is very similar between the two approaches that use ensembles, the one that uses RF shows larger biases. At the warm tail ($p98$ bias), most models perform well, with the notable exception of DeepESD-STO, which overestimates the 98th percentile with a median bias of approximately $0.5 K$. For annual minimum temperature (TNn bias), the Gaussian diffusion model (ClimDiT) yields the most accurate results. DeepESD-MSE also performs reasonably well in terms of median TNn bias, although it exhibits substantial spatial dispersion across the domain. The rectified flow variants and UNet-MULT show moderate biases, although these remain smaller than those of DeepESD-STO.

Temporal variability, assessed through the standard deviation ratio (Std. ratio), is best reproduced by the rectified flow approach and UNet-MULT, whose values lie closest to the ideal value of 1. The Gaussian diffusion model (ClimDiT) and DeepESD-MSE slightly overestimate the observed temporal variability, whereas DeepESD-STO shows the weakest agreement with the observed standard deviation.

Finally, the probabilistic metrics, CRPS and the Brier Score for cold events ($\text{BS } (>p02)$), highlight the advantages of stochastic generation. The single-realization generative models (ClimDiT and ClimDiT-RF) perform similarly to the deterministic baselines, for which CRPS is mathematically equivalent to MAE. By contrast, the 50-member generative ensembles (ClimDiT-ENS and ClimDiT-RF-ENS) achieve performance comparable to that of the purpose-built probabilistic model DeepESD-STO. By effectively modulating the predicted empirical cumulative distribution function (CDF), these ensembles achieve competitive CRPS and Brier score values. These results indicate that the generative ensembles can provide reliable uncertainty estimates while retaining strong point-wise accuracy.

\subsubsection{Maximum temperature downscaling}

\begin{figure}
    \centering
    \includegraphics[width=\linewidth]{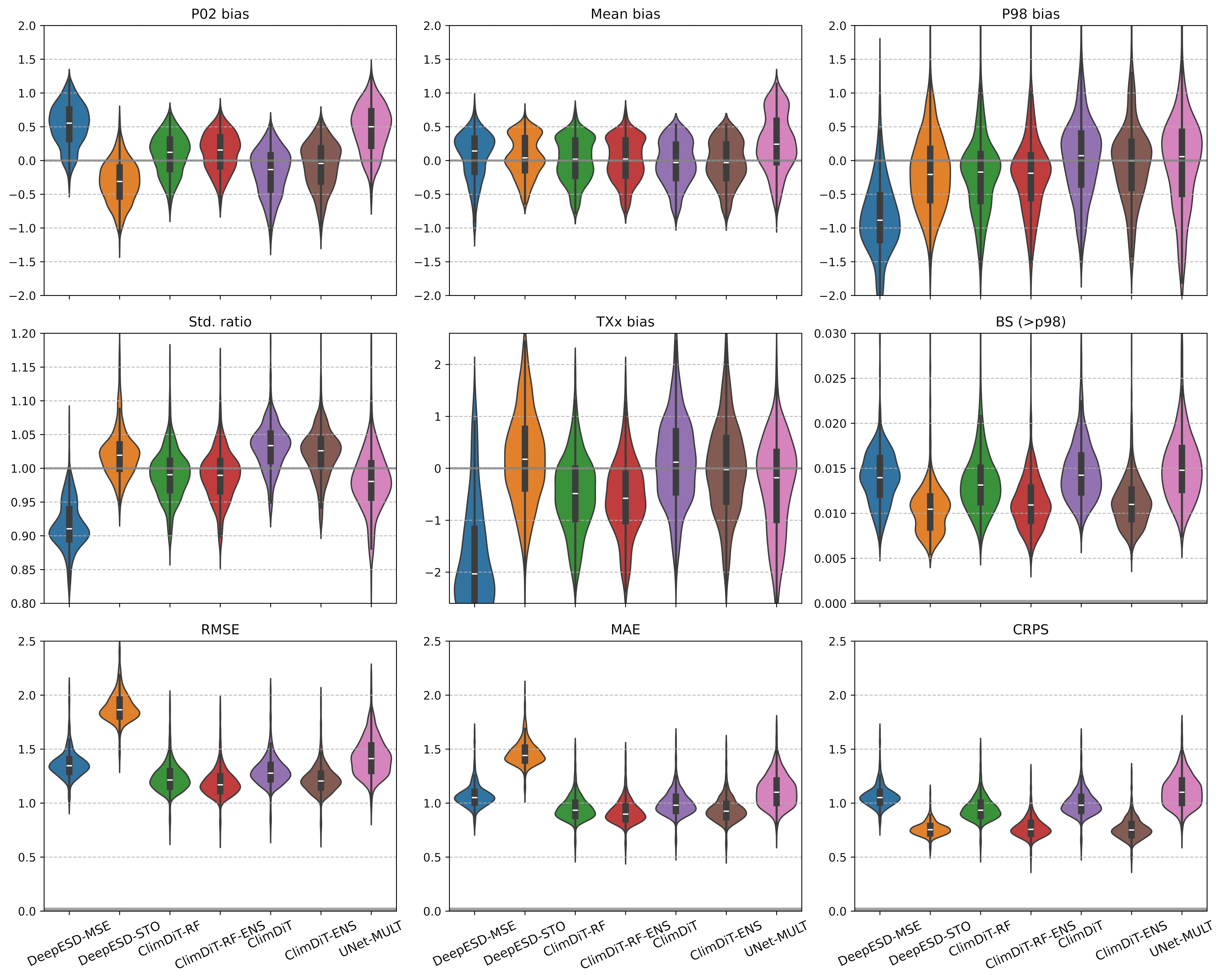}
    \caption{Violin plots of metrics for maximum temperature}
    \label{fig:ViolinMaxTemp}
\end{figure}

Figure \ref{fig:ViolinMaxTemp} displays the spatial distribution of the evaluation metrics for daily maximum temperature ($T_{max}$). Consistent with the results for minimum temperatures, the generative models show robust performance, effectively balancing deterministic accuracy and probabilistic reliability.

Regarding point-wise deterministic errors (RMSE and MAE), the generative frameworks, both ClimDiT and ClimDiT-RF, together with their ensemble variants, achieve the lowest median errors across the domain. They are closely followed by the deterministic baselines, DeepESD-MSE and UNet-MULT Conversely, the stochastic baseline, DeepESD-STO, yields clearly higher deterministic errors, reflecting the expected trade-off associated with its probabilistic formulation.

In terms of climatological biases, the models generally perform well for the mean bias, with the generative models and DeepESD-STO showing the closest agreement with zero, while UNet-MULT exhibits a slight positive bias. At the lower tail of the maximum temperature distribution ($p02$ bias), the generative frameworks present the lowest bias, closely followed by DeepESD-STO. By contrast, both UNet-MULT and DeepESD-MSE tend to overestimate this lower percentile, with spatial medians of approximately $0.5 K$.

The evaluation of warm extremes reveals a substantially larger spatial dispersion across all models than that observed for the cold extremes of $T_{min}$. For the $p98$ bias, the Gaussian diffusion models (ClimDiT and ClimDiT-ENS), together with the deterministic UNet-MULT baseline, exhibit bias distributions centred close to zero. The rectified flow models and DeepESD-STO show a slight underestimation of this upper tail. DeepESD-MSE performs worst in reproducing warm extremes, with a clear underestimation with spatial medians approaching $-1 K$. A similar pattern is observed for annual maximum temperature (TXx bias): ClimDiT, ClimDiT-ENS, UNet-MULT, and DeepESD-STO maintain median biases close to zero, whereas the RF variants show a slight underestimation and DeepESD-MSE exhibits a substantially larger negative bias.

Regarding the standard deviation ratio (Std. ratio), temporal variability is well reproduced by both the generative frameworks and DeepESD-STO. UNet-MULT preserves the median variability effectively but shows larger spatial dispersion, while DeepESD-MSE clearly underestimates the observed temporal standard deviation.

Finally, probabilistic performance is evaluated using the continuous ranked probability score (CRPS) and the Brier score for warm events ($\text{BS } (>p98)$). As expected, the stochastic and generative approaches perform particularly well according to these probabilistic metrics. The 50-member ensembles generated by both ClimDiT and ClimDiT-RF achieve Brier scores that match those of the dedicated stochastic baseline (DeepESD-STO). A similar pattern is observed for CRPS, for which the empirical cumulative distribution function (CDF) derived from the generative ensembles performs comparably to the parametric Gaussian predictive distribution produced by DeepESD-STO. Consistent with the results for the minimum temperature, these findings indicate that the diffusion ensembles achieve a favourable balance between deterministic accuracy and probabilistic performance.

\subsubsection{Precipitation downscaling}\label{sec:PrecResults}

\begin{figure}
    \centering
    \includegraphics[width=\linewidth]{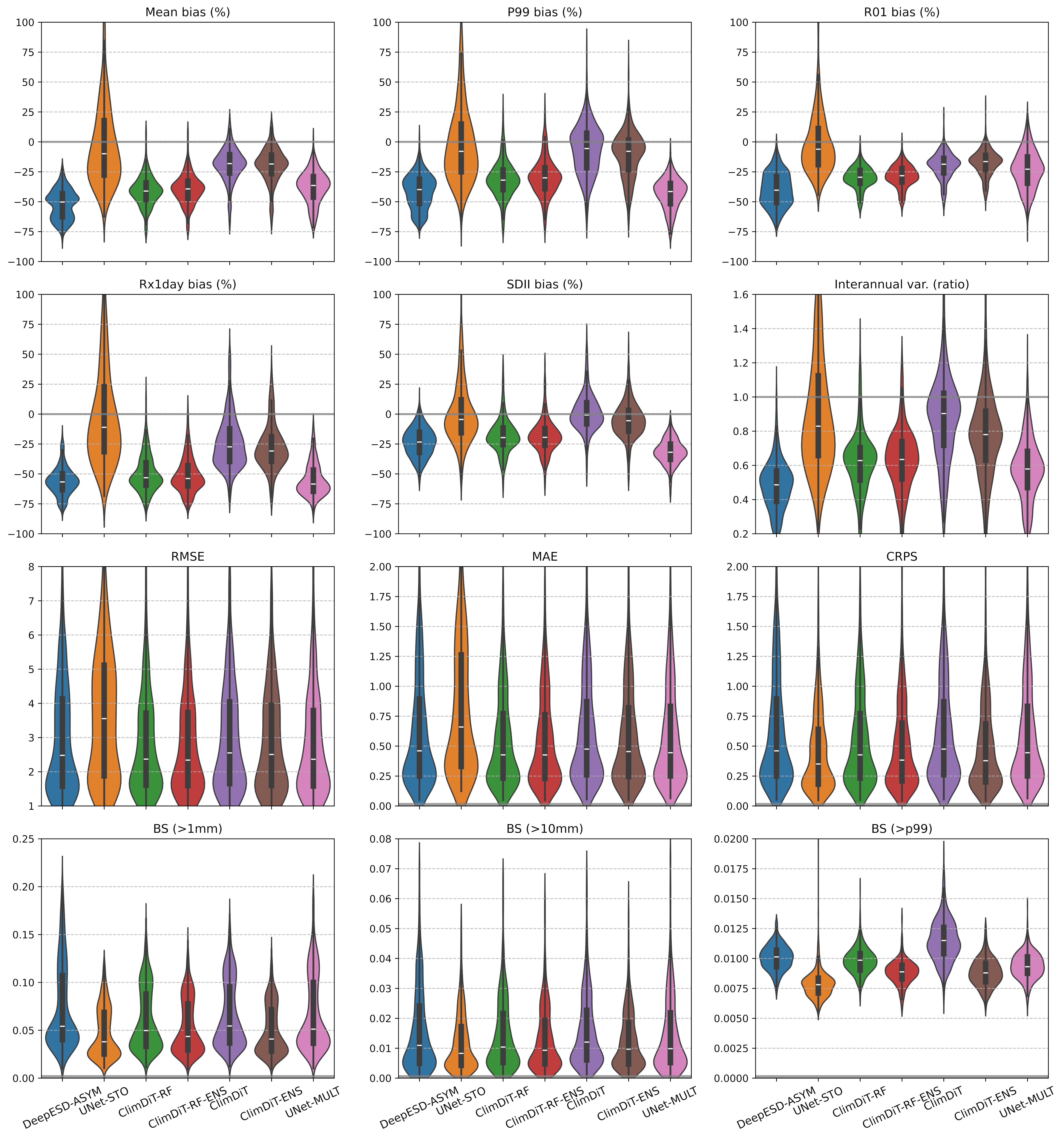}
    \caption{Violin plots of metrics for precipitation}
    \label{fig:ViolinPrec}
\end{figure}

Figure \ref{fig:ViolinPrec} illustrates the spatial distribution of the evaluation metrics for daily accumulated precipitation. 
Regarding deterministic errors (RMSE and MAE), the stochastic baseline (UNet-STO) exhibits the highest error magnitudes and the largest spatial dispersion. This behaviour is consistent with the effects of grid-point-wise stochastic sampling, which can reduce deterministic accuracy. By contrast, the generative models (ClimDiT and ClimDiT-RF) and the deterministic baselines (DeepESD-ASYM and UNet-MULT) achieve comparable performance and yield the lowest deterministic errors.

\begin{figure}
    \centering
    \includegraphics[width=0.9\linewidth]{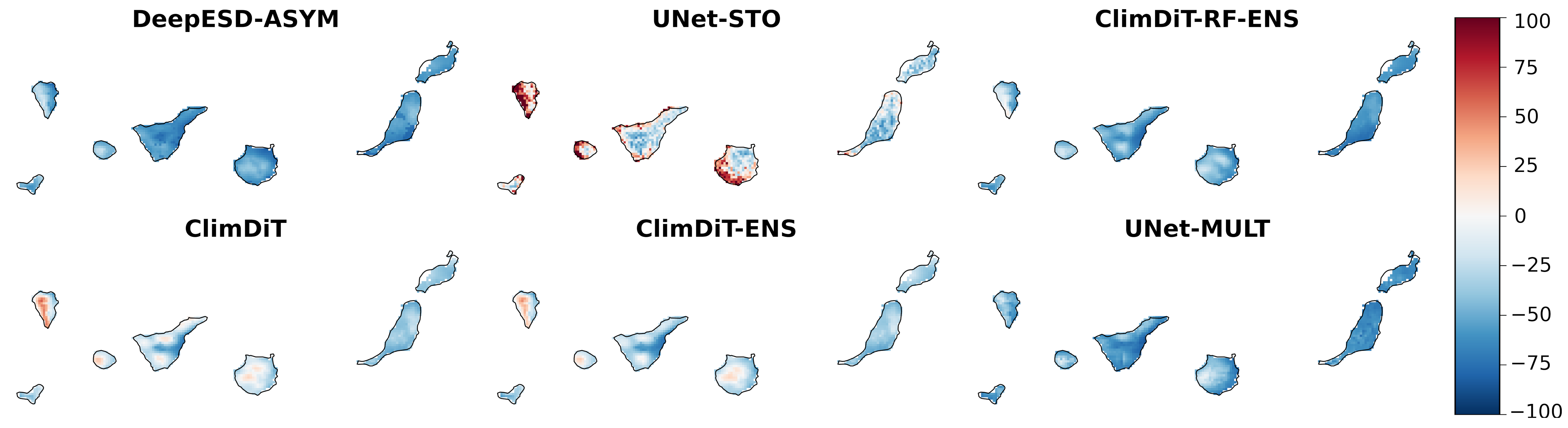}
    \caption{Spatial distribution of the Rx1day bias for daily precipitation.}
    \label{fig:Rx1day_bias_maps}
\end{figure}

For the mean bias, R01 bias (wet-day ratio), and Rx1day bias (maximum daily precipitation), UNet-STO shows median biases close to the ideal value of zero. However, these distributions are characterized by remarkably wide interquartile ranges, indicating substantial spatial variability in model performance. This is particularly evident in the Rx1day maps shown in Figure \ref{fig:Rx1day_bias_maps}, where UNet-STO exhibits an abrupt spatial pattern of adjacent positive and negative biases; consequently, its near-zero spatial median appears to arise partly from compensating local errors. The Gaussian diffusion models (ClimDiT and ClimDiT-ENS) exhibit slight negative biases (underprediction) across these metrics but show considerably narrower spatial distributions, indicating more robust and spatially consistent performance. This behavior is also evident in the Rx1day bias maps, where values remain closer to zero across several islands. By contrast, the rectified flow models, UNet-MULT, and particularly DeepESD-ASYM, show more pronounced underestimation of precipitation across the domain. For the simple daily intensity index (SDII bias), ClimDiT and its 50-member ensemble achieve the lowest overall biases, closely followed by UNet-STO, whereas UNet-MULT exhibits the largest underestimation. These results indicate that, although ClimDiT and ClimDiT-ENS underestimate the frequency of wet days, as reflected by the R01 bias, they reproduce the observed precipitation intensity on wet days reasonably well.

A critical aspect of precipitation downscaling is the representation of extremes. The P99 bias highlights a clear superiority of the gaussian diffusion approach: ClimDiT and its ensemble exhibit bias distributions centred close to zero with relatively limited spatial dispersion. Although UNet-STO also has a median bias close to zero, its spatial variability is considerably larger. The remaining models (ClimDiT-RF, UNet-MULT, and DeepESD-ASYM) systematically underestimate extreme precipitation. A similar pattern is observed for interannual variability, assessed using the interannual variance ratio: UNet-STO, ClimDiT, and ClimDiT-ENS produce values close to the ideal value of 1, whereas the other models tend to underestimate the observed variability.

For the probabilistic metrics (CRPS, BS $>$ 1mm, and BS $>$ 10mm), the differences between the deterministic forecasts, the generative ensembles, and UNet-STO are relatively small. However, for extreme events (BS $>$ p99), the benefits of probabilistic modeling become more evident. UNet-STO achieves the best score for this extreme threshold, followed by the generative ensembles. Nevertheless, although ensembles based on generative models substantially improve probabilistic performance for temperature, their improvement for precipitation is limited, as the deterministic UNet-MULT model performs only slightly worse for this metric. This result reflects the difficulty of accurately representing the complex and highly skewed distribution of precipitation.

\subsubsection{Multivariate performance}

To evaluate multivariate consistency across the three target variables ($T_{max}$, $T_{min}$, and log1p(Prec)), three complementary scores are analyzed: the Euclidean score (EucS), the energy score (ES), and the variogram score (VS). The Euclidean score is used as a deterministic multivariate error metric, whereas ES and VS evaluate the quality of probabilistic multivariate predictions. In this evaluation, ``DET'' denoites the aggregation of the univariate deterministic models (DeepESD-MSE for temperatures and DeepESD-ASYM for precipitation), whereas ``STO'' denotes the aggregation of the univariate stochastic models (DeepESD-STO and UNet-STO).

\begin{figure}
    \centering
    \includegraphics[width=\linewidth]{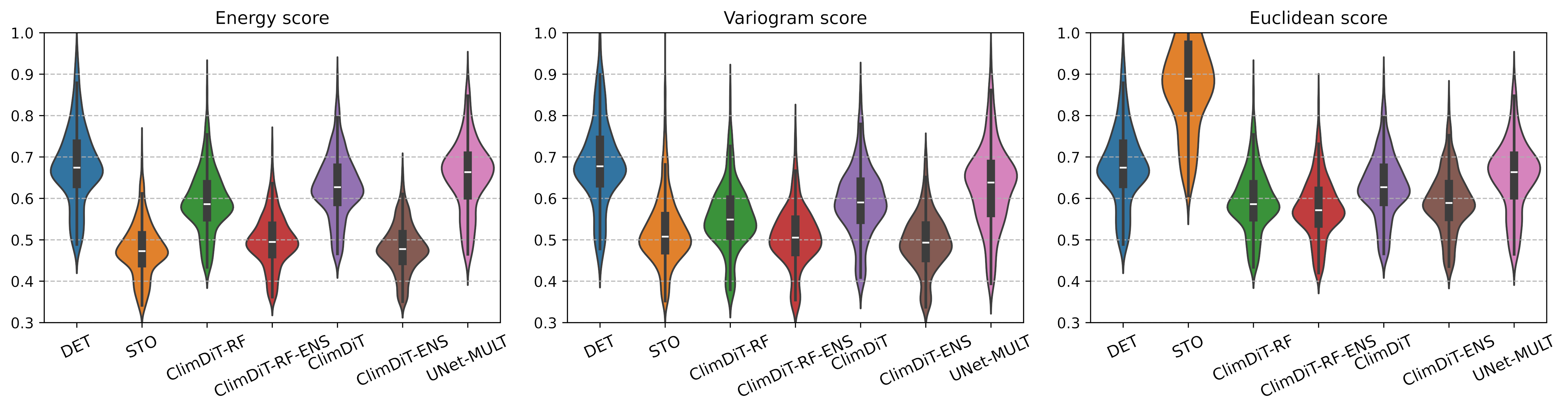}
    \caption{Energy score (ES), variogram score (VS), and multivariate Euclidean score.}
    \label{fig:MultiVarMetrics}
\end{figure}

Figure \ref{fig:MultiVarMetrics} presents the spatial distribution of these multivariate scores. For the EucS, the generative models achieve the lowest errors, with only a modest additional improvement when moving from a single realization to the corresponding ensemble. By contrast, the aggregated stochastic baseline (STO) performs clearly worse, suggesting that its sampling strategy is less effective when the three variables are evaluated jointly.

The ES, which jointly assesses multivariate calibration and sharpness, shows that STO achieves values comparable to those of the generative ensembles, suggesting that the aggregated univariate stochastic models provide competitive uncertainty estimation. By contrast, the degenerate ensembles ($M=1$), corresponding to DET, UNet-MULT, and the single-realization generative models, show poorer ES because they cannot effectively represent predictive uncertainty. Within this group, ClimDiT and, especially, ClimDiT-RF still provide the best performance.

The VS focuses more directly on the dependence structure and the consistency of pairwise increments between variables, making it a useful metric for assessing multivariate predictive performance. For this metric, STO also achieves values comparable to those of the generative ensembles. This result suggests that the stochastic aggregation captures part of the joint variability measured by VS despite being built from univariate models. It also indicates that, according to this dependency-oriented probabilistic score, the explicitly multivariate models do not provide clear advantage over STO. As observed for ES, ClimDiT and ClimDiT-RF are the best-performing models among the degenerate ensembles, indicating better representation of these intervariable relationships within this group. 

Overall, these results indicate that the generative models provide more accurate multivariate point predictions, while their ensemble variants mainly improve the representation of predictive uncertainty rather than substantially reducing deterministic multivariate error.

\subsection{The impact of the latent space}

This subsection analyzes how the latent representation affects the behavior of the proposed downscaling framework from three complementary perspectives. First, it compares VAE-based diffusion with direct pixel-space diffusion to assess the role of latent compression in representing high-resolution climate fields. Second, it investigates the effect of the latent space on optimization stability and sampling efficiency, particularly when reducing the number of DDPM sampling steps. Finally, it evaluates whether the same VAE representation can be transferred effectively to deterministic CNN baselines, thereby clarifying whether its benefits are specific to the diffusion-transformer formulation or are more broadly applicable across model families.

\subsubsection{VAE versus pixel-space diffusion}

Training diffusion transformers directly in the high-resolution pixel space, without the spatial compression provided by a variational autoencoder (VAE), poses substantial computational and optimization challenges. To empirically assess the benefit of incorporating the VAE in the proposed generative frameworks, alternative ClimDiT configurations operating directly on the uncompressed spatial grids were evaluated. To compensate for the quadratic increase in sequence length, an initial configuration used a transformer patch size of $8\times8$. A second configuration used a reduced patch size of $4\times4$ to better preserve local spatial details, despite the substantially higher computational cost.

\begin{figure}
    \centering
    \includegraphics[width=\linewidth]{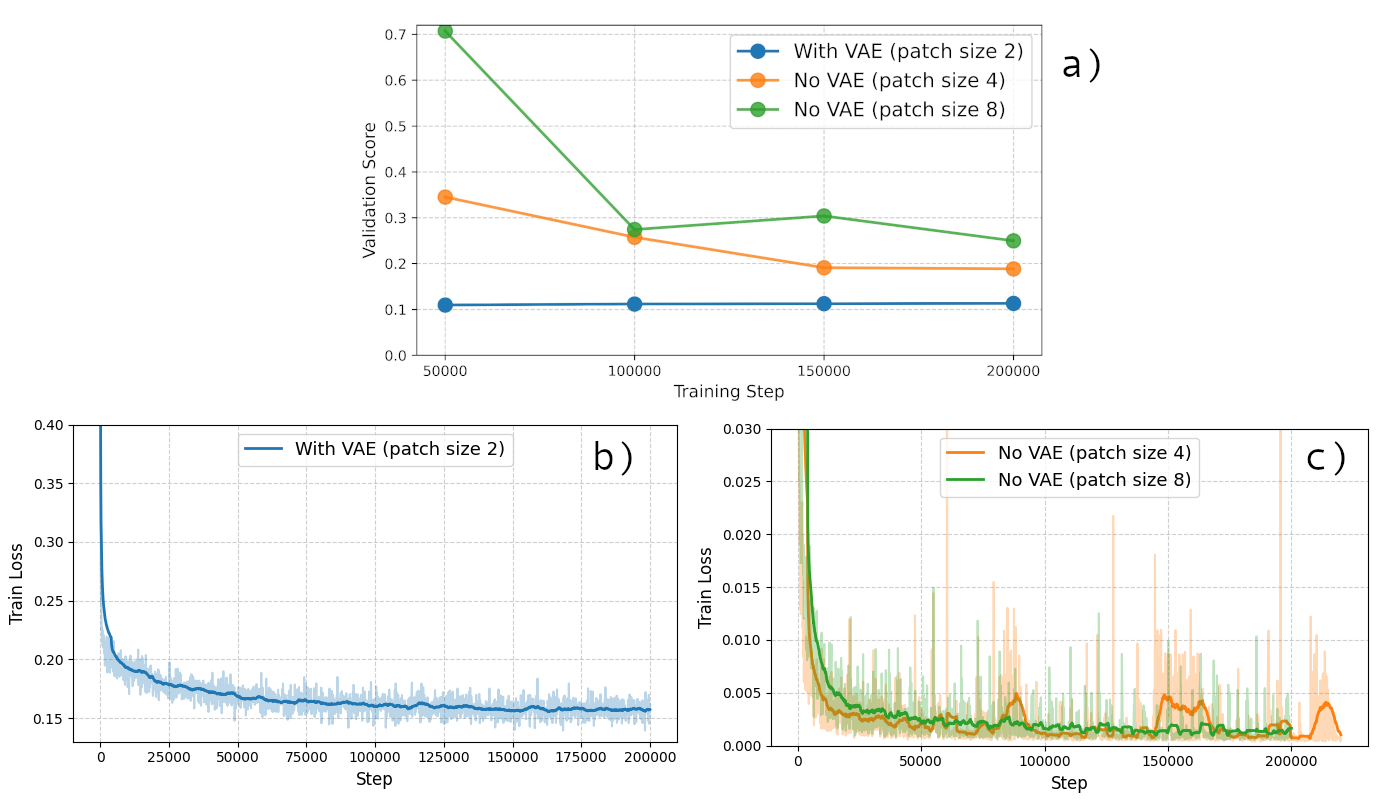}
    \caption{Comparative analysis of ClimDiT training metrics. (a) Evolution of composite validation scores, showing consistent performance across configurations. (b) and (c) show training loss for models with and without a VAE, respectively. Note that while validation scores are directly comparable (computed in uncompressed space), the training loss magnitudes differ because (b) is computed in the latent space while (c) is evaluated in the original pixel space.}
    \label{fig:Training}
\end{figure}

Figure \ref{fig:Training}a shows the evolution of the validation score during training. The VAE-based model achieves a substantially lower and more stable validation score from the early stages of training. By contrast, the No-VAE configurations exhibit markedly poorer performance. Reducing the patch size from $8\times8$ to $4\times4$ slightly improves the validation score by preserving finer spatial details, but it also drastically increases computational cost and still performs considerably worse than the VAE-based configuration. 

These results highlight the challenges of training diffusion transformers directly in pixel space. Operating in the pixel space imposes a severe trade-off between sequence length, i.e. the number of tokens processed, and spatial expressiveness. Maintaining a small patch size (e.g., $4\times4$) to preserve topographic details leads to a quadratic $\mathcal{O}(N^2)$ increase in the transformer sequence length, making it computationally expensive and overwhelming the global attention mechanism. Conversely, increasing the patch size (e.g., $8\times8$) to match the latent sequence length results in a  loss of high-frequency information, since a simple linear patch embedding lacks the deep expressive capacity to decode microclimatic complexity from raw pixels. The VAE alleviates this bottleneck by assigning the reconstruction of local spatial structures to a deep nonlinear convolutional decoder, allowing ClimDiT to focus primarily on the larger-scale dependencies represented in latent space. Without this compression, the attention mechanism must process a much larger set of relatively low-level pixel tokens, resulting in poorer optimization and generalization.

\subsubsection{Optimization stability and sampling efficiency}

Furthermore, operating directly in the pixel space substantially alters optimization dynamics. As shown in Figure \ref{fig:Training}b, the training loss of the VAE-based model converges smoothly and stabilizes after approximately 50,000 steps, indicating more stable convergence. By contrast, Figure \ref{fig:Training}c reveals that both No-VAE configurations (patch sizes $4\times4$ and $8\times8$) exhibit markedly noisier loss trajectories with recurrent spikes. These results are consistent with the idea that the VAE maps the high-dimensional climate fields onto a smoother and more compact latent representation, thereby simplifying the learning problem for the diffusion model. Without this prior compression, the model must simultaneously represent fine-scale spatial variability and large-scale atmospheric structure directly in pixel space, which may contribute to less stable optimization and poorer convergence.

Additionally, the latent space plays an important role in improving computational efficiency during inference. The VAE-based model generates stable high-fidelity samples using only 100 integration steps, whereas applying the same reduced step count in the pixel space substantially degrades the representation of precipitation, as shown in Appendix A. Temperature fields, which are generally smoother and more spatially continuous, appear to be less sensitive to coarser temporal discretization. Precipitation, by contrast, is highly intermittent and contains stronger fine-scale spatial variability, making it more difficult to reproduce accurately with a reduced number of sampling steps. The KL-regularized VAE maps these  physically distinct variables into a compact, approximately Gaussian latent representation, which simplifies the generative process and enables robust sampling with approximately one order of magnitude fewer steps.

Consequently, the VAE is not merely a computational shortcut, but an important architectural component of the proposed framework. By compressing spatial fields into a regularized latent representation, it stabilizes training dynamics and enables high-fidelity multivariate downscaling at a manageable computational cost.

\subsubsection{Transfer to deterministic baselines}

Additionally, to assess whether the benefits of the VAE extend to more conventional regression models, DeepESD-MULT was trained using VAE latent representations instead of the original high-resolution target fields. As shown in Appendix A, this VAE-based DeepESD-MULT configuration did not improve performance relative to the standard DeepESD-MULT setup, suggesting that the benefits observed here are primarily associated with latent diffusion models. An analogous VAE-based extension was not tested for UNet-MULT because it would require substantial architectural modifications: the coarse ERA5 predictors are defined on a $32\times32$ grid, whereas the VAE-compressed ROCIO-CAN target has spatial dimensions $9\times24$. Reconciling these resolutions would require important architectural restructuring of the encoder-decoder pathway and could compromise the characteristic U-shaped architecture that contributes to the effectiveness of U-Net models. 

\subsection{Computational efficiency and memory utilization}

Beyond predictive skill, the practical usefulness of a generative downscaling model depends on whether training and, especially, inference remain computationally affordable for a high-resolution regional domain. This section quantifies the efficiency gains obtained over the Canary Islands from the lighter ClimDiT backbone described in Subsection \nameref{sec:Arch} and the reduced 100-step sampling configuration used in the Gaussian diffusion formulation. The computational requirements of the proposed models are also compared with those of the benchmark models.

For the Gaussian diffusion model, the 200,000-step training run using the previous architecture \citep{EsquivelGonzalez2026climdit} required 286 minutes on four parallel NVIDIA A100 GPUs, with a peak VRAM usage of approximately 8.5 GB per GPU. By contrast, the optimized architecture completes the same training run in 236 minutes, corresponding to a 17\% reduction in total training time. Its peak VRAM usage also decreases to approximately 5.5 GB per GPU, reflecting the smaller parameter count and reduced memory requirements of the updated ClimDiT backbone. In the case of non-generative models (benchmarks), all of them require less than one hour of training on a single NVIDIA A-100 GPU and use less than 4 GB of VRAM, highlighting their lower computational resource requirements.

The advantages of the regularized latent space and architectural optimization are even more pronounced during the inference phase. Evaluating the full 10-year test set, comprising approximately 3,650 days, over the Canary Islands requires 18 minutes on a single NVIDIA A100 GPU with a batch size of 100  when using the traditional 1000-step DDPM sampling trajectory. This time is reduced to 5 minutes when inference is parallelized across four GPUs. By reducing the sampling trajectory to 100 denoising steps, the total inference time for the test set decreases to approximately 2 minutes on a single GPU, corresponding to an 8.6-fold speed-up and a runtime comparable to that of the 100-step rectified flow formulation.

To evaluate applicability on consumer-grade hardware, some benchmark models were executed on a NVIDIA GeForce RTX 4060 GPU with 8 GB of VRAM using a batch size of 32. Table \ref{tab:inference_performance} summarizes the total inference time (scaled to simulate the full diffusion denoising process plus the VAE decoder pass) and peak VRAM allocation for the Canary Islands domain. 

\begin{table}
\begin{center}
\begin{tabular}{lcc}
\topline
\textbf{Model}
& \textbf{Inference time (s)}
& \textbf{Peak VRAM (MB)} \\
\midline
DeepESD-MULT
& 0.01
& 1,662 \\

UNet-MULT
& 0.03
& 1,101 \\
\midline
Previous ClimDiT (\(3 \times 3\) patches)
& 5.2
& 2,233 \\

ClimDiT (\(2 \times 2\) patches)
& 4.4
& 2,125 \\
\botline
\end{tabular}
\caption{Inference performance on an NVIDIA GeForce RTX 4060 GPU with 8 GB of VRAM. The reported values correspond to the generation of a 32-day batch and are extrapolated to represent the full sampling trajectory. The previous ClimDiT configuration is taken from \citet{EsquivelGonzalez2026climdit}.}
\label{tab:inference_performance}
\end{center}
\end{table}

As shown in Table \ref{tab:inference_performance}, the CNN baselines (DeepESD-MULT and UNet-MULT) are substantially faster because they require only a single forward pass, with inference times of less than one second. Nevertheless, the approximately 4.4 s required by the optimized ClimDiT configuration with $2\times2$ patches remains computationally practical for the evaluated setting. When contextualized within the broader field of regional climate modeling, generative statistical downscaling is considerably less computationally demanding than dynamical downscaling with regional climate models (RCMs), which typically require weeks or months on supercomputing clusters to simulate equivalent periods. Thus, the additional inference cost of the iterative diffusion process relative to standard CNNs therefore represents a manageable trade-off for the improvements observed in spatial realism, spectral power recovery, and multivariate probabilistic performance.

For regional downscaling domains such as the Canary Islands, the combination of a $\times8$ VAE spatial compression and a $2\times2$ token patch size represents an effective trade-off to resolve localized microclimatic features and limiting the computational cost associated with the number of attention patches. For future applications involving substantially larger high-resolution grids, this attention bottleneck could be alleviated by using larger patch sizes or incorporating advanced autoencoders with higher spatial downsampling factors.


\section{Conclusions}\label{sec:Conclusions}

This study assessed latent generative models for multivariate statistical downscaling over the Canary Islands, focusing on daily maximum temperature, daily minimum temperature, and precipitation. Beyond extending ClimDiT to a complex island domain, the work introduced several methodological changes: a lighter ClimDiT backbone with fewer trainable parameters, a rectified flow variant of the original Gaussian diffusion framework, and a reduced 100-step DDPM sampling configuration. Additionally, it introduced a composite validation strategy for checkpoint selection. By combining RMSE with extreme-percentile errors, this criterion helped limit the degradation of extreme-value performance during training. The experiments demonstrated its utility not only for the proposed generative frameworks but also for multivariate benchmarks, for which it improved model selection relative to RMSE-only checkpointing. The use of a comprehensive evaluation, allowed the models to be assessed not only in terms of point-wise errors but also in terms of their ability to reproduce spatial structure, predictive uncertainty, and intervariable dependence.

The results indicate that ClimDiT, the latent diffusion model, is a competitive alternative to established CNN-based downscaling approaches. For temperature, the generative models achieve deterministic errors and climatological biases comparable to the best deterministic baselines, while their ensemble variants provide probabilistic performance comparable to that of the dedicated stochastic models. For precipitation, the differences between the approaches are more pronounced. The deterministic baselines tend to produce overly smooth fields and underestimate spectral power, whereas UNet-STO introduces excessive fine-scale variability, exhibits larger spatial dispersion, and yields poorer deterministic performance. ClimDiT provides a more balanced behaviour: it preserves substantially more spatial variance, reproduces the observed precipitation spectrum more closely, and reduces the systematic underestimation of heavy precipitation relative to most of the alternative models.

The comparison with rectified flow shows that not all latent generative formulations perform equally. ClimDiT-RF and its ensemble variant (ClimDiT-RF-ENS) perform well across all metrics, particularly in RMSE and MAE, but they show a higher tendency than the Gaussian diffusion models to underestimate precipitation amounts, spectral power, and some extreme percentiles. Rectified flow also offers the potential for further reductions in sampling cost through ReFlow-style distillation. However, because such methods are primarily designed to reduce the number of sampling steps rather than improve the learned distribution, they are unlikely to address the main performance limitations observed for ClimDiT-RF in this downscaling task.

The multivariate evaluation shows that ClimDiT and ClimDiT-RF provide the best Euclidean score distributions, indicating more accurate multivariate point predictions, whereas the aggregated stochastic baseline (STO) performs substantially worse for this deterministic joint metric. However, for the probabilistic multivariate scores, the energy score and variogram score, STO remains highly competitive, achieving values comparable to those of ClimDiT-ENS and ClimDiT-RF-ENS. Thus, the generative ensembles improve the representation of predictive uncertainty, but they do not clearly outperform the aggregation of independently trained univariate stochastic models according to these two scores. 

The ablation experiments further indicate that these results depend strongly on the latent representation. The VAE acts not only as a compression module but also as a regularizing component that stabilizes diffusion-transformer training, preserves spatial and spectral structure, and enables robust 100-step DDPM sampling. Together with the lighter ClimDiT backbone, this reduced-step configuration substantially lowers inference time and memory requirements while retaining the main performance advantages of latent diffusion. By contrast, using the same VAE representation with DeepESD-MULT does not improve performance relative to the standard deterministic configuration, suggesting that its benefits are most relevant when coupled with probabilistic latent generative models.

Overall, the findings support latent Gaussian diffusion models as a promising framework for multivariate climate downscaling in complex high-resolution domains. Their main strengths include the preservation of fine-scale precipitation structure, competitive deterministic accuracy, strong multivariate consistency, and operationally feasible inference with the optimized 100-step sampling configuration. Future work should evaluate latent generative downscaling models within standardized benchmarks \citep{rampal2026cordexmlbench} that enable direct comparison with other deterministic, probabilistic, and generative approaches. These comparisons should cover different settings, including perfect-prognosis experiments, climate-model emulation, and future climate projections. Further research should also assess transferability across regions and climates, explore more efficient samplers and loss formulations, and examine the reliability of the generated distributions under non-stationary climate conditions.


\clearpage
\acknowledgments
This work was supported by the PLANCLIMAC2 Project (1/MAC/2/2.4/0006), financed through the European Union’s INTERREG MAC 2021–2027 Program. Additional support was provided by the Government of the Canary Islands, through the Consejería de Transición Ecológica y Energía, as part of the cooperation agreement signed on November 8, 2024. The methods developed in this work will be used at higher spatial resolution in archipelagos within the CORDEX Flagship Pilot Study: Macaronesian Archipelagos: Convection Permitting projections focused on island processes (FPS-I-Mac). 
The authors have no relevant financial or non-financial interests to disclose.

%
%
\datastatement
All datasets used in this study are publicly available. The ERA5 reanalysis data can be obtained from the Copernicus Climate Data Store (CDS) upon registration and acceptance of the license agreement (\url{https://doi.org/10.24381/cds.50314f4c}). The ROCIO+\_CAN observational dataset is hosted by AEMET and can be accessed through its website (\url{https://www.aemet.es/en/serviciosclimaticos/cambio_climat/datos_diarios?w=2}). All code used to reproduce this study is openly available on GitHub (\url{https://github.com/marcosesgonz/ClimDiT-downscaling}).








%



\appendix[A] 

\appendixtitle{Ablation study}
\label{sec:appendix_ablation}

This appendix summarizes the auxiliary experiments conducted to support the model-selection decisions described in the main text. The ablation study focuses on three methodological choices: the use of the composite validation score (CS) for checkpoint selection, the effect of coupling the VAE latent representation with different model families, and the sensitivity of ClimDiT to reducing the number of DDPM sampling steps from 1,000 to 100. Table \ref{tab:ablation_spatial_medians} reports the spatial-median results obtained in these experiments across a wide set of evaluation metrics.

\begin{table}[!htbp]
\begin{center}
\caption{Spatial-median ablation results for the main univariate, multivariate,
and spatial evaluation metrics for the test period. CS denotes checkpoint selection based on the
composite validation score (Eqs. 9--11). The ClimDiT-CS column corresponds to
the no-VAE pixel-space configuration. For ClimDiT, paired values report results
obtained using 1,000 and 100 sampling steps, respectively. Lower values are better for RMSE, MAE, CRPS, BS, ES, and VS. Higher values are better for SSIM. Variability ratios are optimal near 1.}
\label{tab:ablation_spatial_medians}

\scriptsize
\setlength{\tabcolsep}{2.5pt}

\begin{tabular}{lccccccc}
\topline
&
\multicolumn{3}{c}{\textbf{DeepESD-MULT}} &
\multicolumn{2}{c}{\textbf{UNet-MULT}} &
\multicolumn{2}{c}{\textbf{ClimDiT} (1000/100 steps)} \\
\midline
\textbf{Metric}
& \textbf{VAE + CS}
& \textbf{CS}
& \textbf{No CS}
& \textbf{CS}
& \textbf{No CS}
& \textbf{VAE + CS}
& \textbf{CS} \\
\hline
\multicolumn{8}{c}{\textbf{Maximum temperature}} \\
\hline
RMSE
& 1.33 & 1.40 & 1.39 & 1.41 & 1.37 & 1.29/1.28 & 1.90/1.96 \\
MAE
& 1.03 & 1.10 & 1.09 & 1.10 & 1.06 & 0.99/0.98 & 1.49/1.53 \\
Mean bias
& 0.07 & 0.00 & -0.06 & 0.24 & 0.04 & -0.03/-0.03 & 0.37/0.43 \\
\(P02\) bias
& 0.37 & 0.23 & 0.33 & 0.50 & 0.60 & -0.12/-0.13 & -0.07/-0.01 \\
\(P98\) bias
& -0.65 & -0.72 & -0.87 & 0.06 & -0.81 & 0.05/0.07 & -0.15/0.12 \\
Std. ratio
& 0.95 & 0.96 & 0.94 & 0.98 & 0.93 & 1.03/1.03 & 1.02/1.03 \\
TXx bias
& -1.56 & -1.65 & -1.84 & -0.18 & -1.33 & 0.17/0.12 & -0.53/-0.08 \\
\(\mathrm{BS}(>P98)\)
& 0.01 & 0.01 & 0.02 & 0.01 & 0.01 & 0.01/0.01 & 0.02/0.02 \\
CRPS
& 1.03 & 1.10 & 1.09 & 1.10 & 1.06 & 0.99/0.98 & 1.49/1.53 \\
SSIM
& 0.95 & 0.94 & 0.94 & 0.94 & 0.93 & 0.95/0.95 & 0.81/0.79 \\
\hline
\multicolumn{8}{c}{\textbf{Minimum temperature}} \\
\hline
RMSE
& 1.09 & 1.14 & 1.12 & 1.11 & 1.09 & 1.06/1.06 & 1.65/1.67 \\
MAE
& 0.85 & 0.90 & 0.88 & 0.87 & 0.85 & 0.83/0.83 & 1.31/1.33 \\
Mean bias
& -0.06 & -0.04 & 0.01 & 0.04 & -0.08 & -0.14/-0.15 & -0.23/-0.10 \\
\(P02\) bias
& 0.44 & 0.36 & 0.52 & 0.32 & 0.55 & -0.14/-0.14 & 0.00/0.08 \\
\(P98\) bias
& -0.20 & -0.11 & -0.21 & -0.07 & -0.67 & 0.02/0.02 & 0.05/0.36 \\
Std. ratio
& 0.96 & 0.99 & 0.97 & 0.99 & 0.92 & 1.04/1.04 & 1.04/1.05 \\
TNn bias
& 0.76 & 0.69 & 1.00 & 0.67 & 0.96 & 0.07/0.06 & 0.15/0.19 \\
\(\mathrm{BS}(<P02)\)
& 0.02 & 0.03 & 0.02 & 0.02 & 0.02 & 0.03/0.03 & 0.03/0.03 \\
CRPS
& 0.85 & 0.90 & 0.88 & 0.87 & 0.85 & 0.83/0.83 & 1.31/1.33 \\
SSIM
& 0.96 & 0.95 & 0.96 & 0.95 & 0.95 & 0.96/0.96 & 0.83/0.81 \\
\hline
\multicolumn{8}{c}{\textbf{Precipitation}} \\
\hline
RMSE
& 2.43 & 2.44 & 2.48 & 2.36 & 2.37 & 2.65/2.55 & 3.13/2.84 \\
MAE
& 0.44 & 0.53 & 0.51 & 0.44 & 0.46 & 0.48/0.48 & 0.73/0.66 \\
Mean bias (\%)
& -49.39 & -30.33 & -49.32 & -36.39 & -41.94
& -19.24/-18.12 & -59.74/-88.23 \\
\(P99\) bias (\%)
& -57.46 & -53.84 & -62.42 & -41.48 & -52.35
& -6.56/-5.57 & -44.73/-65.45 \\
R01 bias (\%)
& -37.79 & -4.47 & -25.84 & -22.83 & -35.35
& -19.41/-18.66 & -4.73/-38.72 \\
Rx1day bias (\%)
& -64.81 & -63.54 & -71.10 & -58.41 & -66.94
& -24.42/-27.35 & -43.88/-62.10 \\
SDII bias (\%)
& -39.93 & -48.04 & -51.28 & -31.84 & -36.94
& -0.84/-0.80 & -34.21/-44.64 \\
IAV ratio
& 0.44 & 0.55 & 0.46 & 0.58 & 0.52 & 0.82/0.90 & 0.66/0.37 \\
\(\mathrm{BS}(>1\,\mathrm{mm})\)
& 0.05 & 0.05 & 0.05 & 0.05 & 0.05 & 0.05/0.05 & 0.09/0.07 \\
\(\mathrm{BS}(>10\,\mathrm{mm})\)
& 0.01 & 0.01 & 0.01 & 0.01 & 0.01 & 0.01/0.01 & 0.01/0.01 \\
\(\mathrm{BS}(>P99)\)
& 0.01 & 0.01 & 0.01 & 0.01 & 0.01 & 0.01/0.01 & 0.01/0.01 \\
CRPS
& 0.44 & 0.53 & 0.51 & 0.44 & 0.46 & 0.48/0.48 & 0.73/0.66 \\
SSIM, \(\log(1+\mathrm{Prec})\)
& 0.71 & 0.39 & 0.39 & 0.71 & 0.59 & 0.78/0.78 & 0.15/0.11 \\
\botline
\end{tabular}
\end{center}
\end{table}

The results first show that the composite validation score is also beneficial for the CNN baselines. Compared with the no-CS configuration, checkpoint selection based on CS improves several precipitation diagnostics, particularly the bias metrics and interannual variability, although some aspects of precipitation remain underestimated. CS also improves several temperature diagnostics, especially those related to extreme-value biases, and leads to better spatial-structure metrics for UNet-MULT.

The VAE ablation leads to a more model-dependent conclusion. For DeepESD-MULT, training in the VAE latent space improves some diagnostics, including precipitation MAE and SSIM, but degrades others, such as mean bias, wet-day frequency, and interannual variability. Thus, the VAE should not be interpreted as a universally beneficial preprocessing step for deterministic CNNs. By contrast, the VAE plays a much more important role in ClimDiT. With the latent VAE representation, ClimDiT achieves the lowest temperature errors and  the highest spatial coherence among the ablation variants. It also provides the best precipitation performance for extreme-percentile bias, SDII bias, interannual variability, and SSIM. Removing the VAE substantially degrades the diffusion model, particularly for precipitation, for which SSIM drops sharply to values close to zero.

The paired ClimDiT results further show that the latent configuration is robust to accelerated sampling. Reducing the number of latent ClimDiT sampling steps from 1,000 to 100 leaves the temperature metrics essentially unchanged and even slightly improves several precipitation diagnostics, including RMSE, mean bias, P99 bias, and interannual variability ratio. By contrast, the no-VAE ClimDiT configuration is considerably more sensitive to the same reduction. It should be noted that the model used is designed to be used in conjunction with the VAE, and it would probably be necessary to redesign the model to operate within the original space. As shown in Table \ref{tab:ablation_spatial_medians}, precipitation biases and the interannual variability ratio deteriorate when the number of sampling steps is reduced. This contrast supports the interpretation that the VAE latent space acts not only as a spatial compression mechanism but also as a regularized representation that enables more robust DDPM sampling with larger step intervals, particularly for intermittent precipitation fields.

\begin{figure}[htbp]
    \centering
    \includegraphics[width=\textwidth]{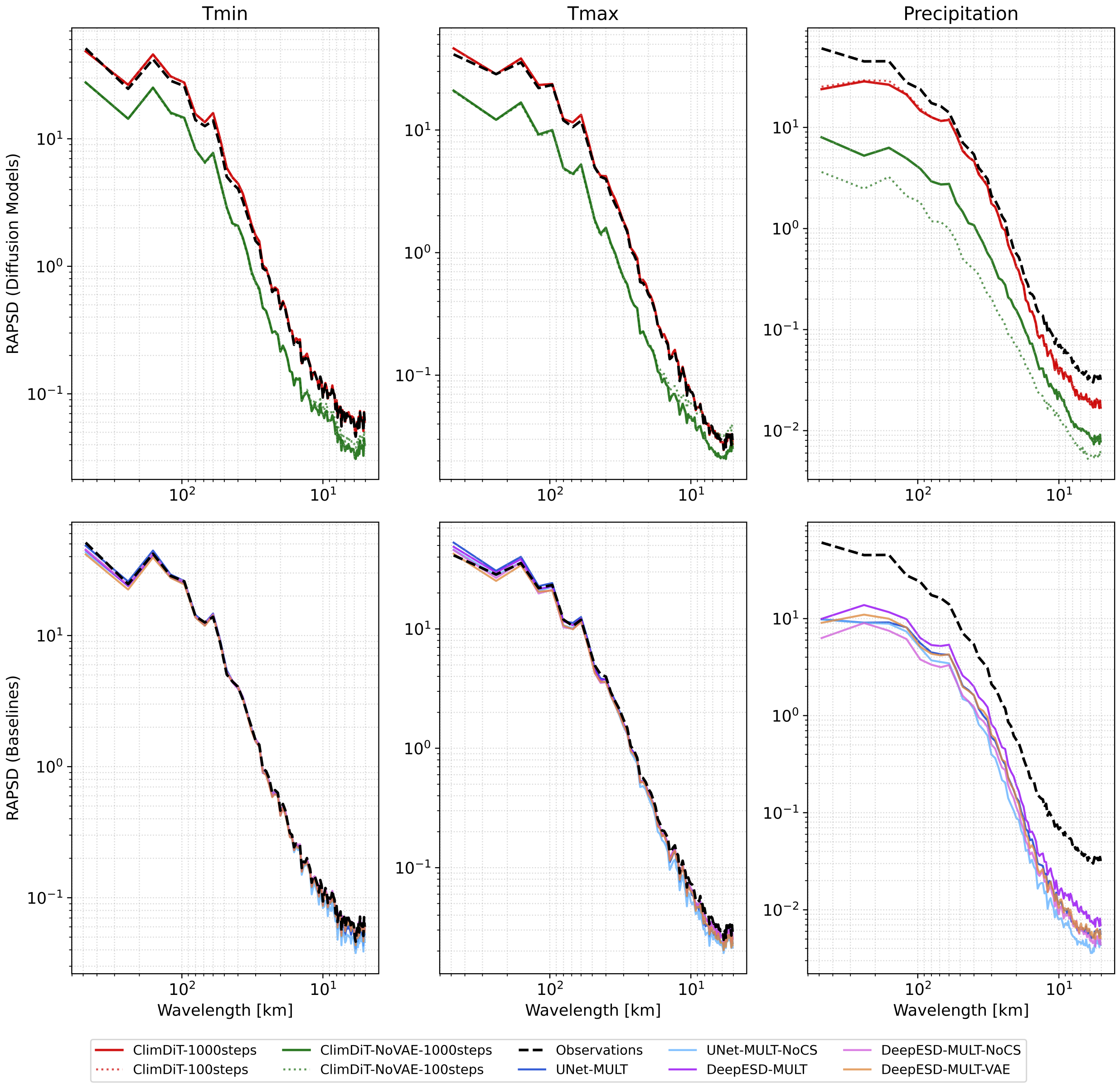}
    \caption{Radially averaged power spectral density (RAPSD) for the ablation experiments. The top row shows the ClimDiT configurations, and the bottom row shows the CNN benchmark configurations. Columns correspond to the target variables.}
    \label{fig:ablation_rapsd_lowres}
\end{figure}

The RAPSD analysis in Figure \ref{fig:ablation_rapsd_lowres} further supports these conclusions. For temperature, the latent ClimDiT configurations closely reproduce the observed spectra using both 1,000 and 100 steps, whereas the no-VAE diffusion variants lose spectral power across a broad range of wavelengths. The CNN baselines also reproduce the temperature spectra well, suggesting that, in this ablation setting, temperature fields are less challenging from a spectral perspective than precipitation fields. For precipitation, however, the differences are considerably more pronounced. Latent ClimDiT preserves substantially more spectral power than its no-VAE counterparts, whereas the 100-step no-VAE variant shows the strongest spectral degradation. The CNN baselines, including the VAE-based DeepESD-MULT configuration, remain well below the observed precipitation spectrum across spatial scales. These results indicate that the spatial and spectral realism achieved by ClimDiT cannot be attributed to VAE compression alone, but instead arises from the interaction between the regularized latent representation and the probabilistic diffusion transformer formulation.

\appendix[B] \label{ap:glossary}

\appendixtitle{Glossary of metrics and formulas}

\subsection{Notation}
Let \(i \in \Omega\) index grid cells (domain), \(t=1,\dots,T\) index time (days), and \(\mathcal{T}_g\) denote the set of days in year \(g\) (with \(G\) total years). Let \(y_t(i)\) denote the observed value at \((t,i)\).
A deterministic model provides \(\hat{y}_t(i)\), and an ensemble model provides values for each member \(x_{t,m}(i)\), \(m=1,\dots,M\).
For precipitation, \(r_t(i)\) denotes daily precipitation; a wet-day threshold is \(r_0\) (\(1\,\mathrm{mm\,day^{-1}}\) in this case). \(T_{\min,t}(i)\) and \(T_{\max,t}(i)\) specify minimum and maximum daily temperature, respectively. Formulas are given per \((t,i)\) or per year as specified.

\paragraph{Averaging operators}
Time average: \(\overline{a}=\frac{1}{T}\sum_{t=1}^T a_t\); sample standard deviation: \(\mathrm{SD}(a)=\sqrt{\frac{1}{T}\sum_{t}(a_t-\overline{a})^2}\). Absolute value \(|\cdot|\), Euclidean norm \(\|\cdot\|_2\).

\subsection{Deterministic metrics}
\begin{itemize}

\item \textbf{Mean bias (MB).}

For temperature (absolute):
\[
\mathrm{MB}(i)=\overline{\hat{y}(i)}-\overline{y(i)}.
\]
For precipitation (relative):
\[
\mathrm{MB}_{\%}(i)=100\cdot
\frac{\overline{\hat{r}(i)}-\overline{r(i)}}{\overline{r(i)}}.
\]

\item \textbf{Percentile bias (PB).} Let \(y_{(p)}(i)\) denote the 
p-th percentile over time of the observed values at grid cell 
$i$. For a given percentile \(p \in (0,1)\):
\[
\mathrm{PB}_p(i) = \hat{y}_{(p)}(i) - y_{(p)}(i),
\]

For precipitation, the \(p = 0.99\) bias is expressed in relative terms:
\[
\mathrm{PB}_{99,\%}(i) = 100 \cdot
\frac{\hat{r}_{(99)}(i) - r_{(99)}(i)}{r_{(99)}(i)}.
\]

\item \textbf{Relative bias for the ratio of wet days (R01 bias).}

Wet-day ratio:
\[
\mathrm{R01}(i)=\frac{1}{T}\sum_t \mathbf{1}\{r_t(i)\ge r_0\}.
\]

Relative bias:
\[
\mathrm{Bias}_{\%}(\mathrm{R01})(i)=100\cdot
\frac{\widehat{\mathrm{R01}}(i)-\mathrm{R01}(i)}{\mathrm{R01}(i)}.
\]

\item \textbf{Relative bias for the Simple Daily Intensity Index (SDII).}
\[
\mathrm{SDII}(i)=
\frac{\sum_t r_t(i)\,\mathbf{1}\{r_t(i)\ge r_0\}}
{\sum_t \mathbf{1}\{r_t(i)\ge r_0\}},
\qquad
\mathrm{Bias}_{\%}(\mathrm{SDII})(i)=100\cdot
\frac{\widehat{\mathrm{SDII}}(i)-\mathrm{SDII}(i)}{\mathrm{SDII}(i)}.
\]

\item \textbf{Relative bias for the climatology of maximum daily precipitation (Rx1day).}
\[
\mathrm{Rx1day}_g(i)=\max_{t\in\mathcal{T}_g} r_t(i),
\quad
\overline{\mathrm{Rx1day}}(i)=\frac{1}{G}\sum_g \mathrm{Rx1day}_g(i).
\]
Relative bias:
\[
100\cdot
\frac{\widehat{\overline{\mathrm{Rx1day}}}(i)-\overline{\mathrm{Rx1day}}(i)}
{\overline{\mathrm{Rx1day}}(i)}.
\]

\item \textbf{Bias for annual temperature extremes (TNn/TXx).}
\[
\mathrm{TNn}_g(i)=\min_{t\in\mathcal{T}_g}T_{\min,t}(i),
\qquad
\mathrm{TXx}_g(i)=\max_{t\in\mathcal{T}_g}T_{\max,t}(i).
\]
Biases (averaged across years):
\[
\overline{\mathrm{TNn}}^{\mathrm{bias}}(i)
=\frac{1}{G}\sum_g(\widehat{\mathrm{TNn}}_g(i)-\mathrm{TNn}_g(i)),
\quad
\text{and analogously for TXx.}
\]

\item \textbf{Root mean square error (RMSE) and mean absolute error (MAE)}.
\[
\mathrm{RMSE}(i)=
\sqrt{\frac{1}{T}\sum_t(\hat{y}_t(i)-y_t(i))^2},
\qquad
\mathrm{MAE}(i)=
\frac{1}{T}\sum_t|\hat{y}_t(i)-y_t(i)|.
\]

\item \textbf{Standard-deviation ratio (temperatures).}
\[
\mathrm{StdRatio}(i)=
\frac{\mathrm{SD}(\hat{y}(i))}{\mathrm{SD}(y(i))}.
\]

\item \textbf{Interannual variability ratio (precipitation).}
For each year \(y\), the annual mean precipitation is
\[
\overline{r}_g(i)=\frac{1}{|\mathcal{T}_g|}\sum_{t\in\mathcal{T}_g} r_t(i).
\]
The ratio of interannual variability is then
\[
\mathrm{IAVR}(i)=
\frac{\mathrm{SD}_g(\widehat{\overline{r}}_g(i))}
{\mathrm{SD}_g(\overline{r}_g(i))}.
\]
\end{itemize}
\subsection{Probabilistic metrics}

\begin{itemize}

\item \textbf{Continuous ranked probability score (CRPS)}.
Let \(F_{t,i}(z) = \Pr(X_t(i) \le z)\) denote the predictive cumulative distribution function (CDF)
for day \(t\) and grid cell \(i\).
\[
\mathrm{CRPS}_{t}(i) =
\int_{-\infty}^{\infty} \big(F_{t,i}(z) - \mathbf{1}\{y_t(i) \le z\}\big)^2\,\mathrm{d}z.
\]
Ensemble estimator (members \(x_{t,m}(i) \sim F_{t,i}\)):
\[
{\mathrm{CRPS}}_{t}(i)
= \frac{1}{M}\sum_{m=1}^M |x_{t,m}(i) - y_t(i)|
- \frac{1}{2M^2}\sum_{m=1}^M\sum_{m'=1}^M |x_{t,m}(i) - x_{t,m'}(i)|.
\]
The CRPS for each grid cell is then averaged over the time period:
\[
\mathrm{CRPS}(i) = \frac{1}{T}\sum_{t=1}^T {\mathrm{CRPS}}_{t}(i).
\]
For deterministic forecasts, \(\mathrm{CRPS} = \mathrm{MAE}\).

\item \textbf{Brier score (BS)}.
For an event \(E\) (e.g., \(T_{\min}<p02\), \(T_{\max}>p98\), \(r\ge u\)) with forecast probability \(p_t(i)\) and outcome \(o_t(i)=\mathbf{1}\{E\text{ occurs}\}\):
\[
\text{BS}(i)=\frac{1}{T}\sum_{t=1}^T\big(p_t(i)-o_t(i)\big)^2.
\]
For an ensemble, \(p_t(i)\) is the fraction of members satisfying \(E\).
\end{itemize}

\subsection{Spatial fidelity metrics}

\begin{itemize}

\item \textbf{Structural similarity index (SSIM)}.
For two fields \(A,B\) on a given day (then averaged over the test period):
\[
\mathrm{SSIM}(A,B)=
\frac{(2\mu_A\mu_B+C_1)(2\sigma_{AB}+C_2)}{(\mu_A^2+\mu_B^2+C_1)(\sigma_A^2+\sigma_B^2+C_2)}.
\]
Here \(\mu_A,\mu_B\) are means, \(\sigma_A^2,\sigma_B^2\) variances and \(\sigma_{AB}\) the covariance over spatial Gaussian windows with $\sigma=1.5$, as implemented in \citet{wang2004}.

\item \textbf{Radially averaged power spectral density (RAPSD)}\citep{Sinclair_2005}.
This method computes the two-dimensional discrete Fourier transform
(DFT) of a field $f(h,w)$ with dimensions $H\times W$ and grid spacing $s$, obtaining
\[
F(u,v) = \frac{1}{HW} \sum_{h=0}^{H-1} \sum_{w=0}^{W-1}
f(h,w)\, e^{-j2\pi\left(\frac{uh}{M}+\frac{vw}{N}\right)}.
\]
The power spectral density is defined as $P(f)=|F(u,v)|^2$. 
Then, the spectrum is averaged in concentric rings. Each ring represents a specific range of frequencies, and the average power within that ring is calculated:
\[
P^r(f_r)=\frac{1}{N_r(f_r)} \sum_{i=1}^{N_r(f_r)} P(f_{r,i}),
\]
where $N_r(f_r)$ denotes the number of discrete frequencies within each ring.
The resulting one-dimensional spectrum $P^r(f_r)$ describes the distribution of
spatial variance across scales, which can also be expressed in terms of the
spatial wavelength $\lambda = s/f_r$. $P^r(f_r)$ is calculated, for a given variable and day, for the prediction and the observation and then averaged over time. In this study, RAPSD is computed using the \texttt{PySTEPS} package. 
\end{itemize}

\subsection{Multivariate metrics}

\paragraph{Standardization and transformation}
Let the multivariate observation at time \(t\) and grid cell \(i\) be
\[
\mathbf{y}_t(i)=\big[T_{\max,t}(i),\,T_{\min,t}(i),\,\log(1+r_t(i))\big]^\top \in \mathbb{R}^d,
\]
with \(d=3\) in our case. For each grid cell \(i\) and component \(k\in\{1,\dots,d\}\), let \(\mu^k(i)\) and \(\sigma^k(i)\) denote the mean and standard deviation of the
observations over the test period. Observations and forecasts
are standarized, before scoring, componentwise as
\[
\tilde{y}_t^k(i) = \frac{\big(y_t^k(i)-\mu^k(i)\big)}{\sigma^k(i)}, \qquad
\tilde{x}_{t,m}^k(i) = \frac{\big(x_{t,m}^k(i)-\mu^k(i)\big)}{\sigma^k(i)},
\]

\begin{itemize}
\item \textbf{Euclidean score (EucS)}. 
The Euclidean Score is used as a deterministic multivariate error metric. For a deterministic prediction \(\tilde{\mathbf{x}}_t(i)\), it is defined at \((t,i)\) as
\[
\mathrm{Euc}_{t}(i)=\big\|\tilde{\mathbf{x}}_t(i)-\tilde{\mathbf{y}}_t(i)\big\|_2.
\]
For ensemble forecasts, the ensemble mean is used as the deterministic prediction before computing the score:
\[
\tilde{\bar{\mathbf{x}}}_t(i)=\frac{1}{M}\sum_{m=1}^{M}\tilde{\mathbf{x}}_{t,m}(i),
\qquad
\mathrm{Euc}_{t}(i)=\big\|\tilde{\bar{\mathbf{x}}}_t(i)-\tilde{\mathbf{y}}_t(i)\big\|_2.
\]
The time-averaged score per grid cell is
\[
\mathrm{Euc}(i)=\frac{1}{T}\sum_{t=1}^{T}\mathrm{Euc}_{t}(i).
\]

\item \textbf{Energy score (ES).}
For a multivariate predictive distribution \(F_{t,i}\) on \(\mathbb{R}^d\) and observation \(\tilde{\mathbf{y}}_t(i)\),
the Energy Score at \((t,i)\) is
\[
\mathrm{ES}_{t}(i)=\mathbb{E}\big\|\mathbf{X}-\tilde{\mathbf{y}}_t(i)\big\|_2
-\tfrac{1}{2}\,\mathbb{E}\big\|\mathbf{X}-\mathbf{X}'\big\|_2,\quad \mathbf{X},\mathbf{X}'\overset{\text{iid}}{\sim} F_{t,i}.
\]
Ensemble estimator:
\[
\widehat{\mathrm{ES}}_{t}(i)=\frac{1}{M}\sum_{m=1}^M \big\|\tilde{\mathbf{x}}_{t,m}(i)-\tilde{\mathbf{y}}_t(i)\big\|_2
-\frac{1}{2M^2}\sum_{m=1}^M\sum_{m'=1}^M \big\|\tilde{\mathbf{x}}_{t,m}(i)-\tilde{\mathbf{x}}_{t,m'}(i)\big\|_2.
\]
Time-averaged per grid cell:
\[
\mathrm{ES}(i)=\frac{1}{T}\sum_{t=1}^{T}\widehat{\mathrm{ES}}_{t}(i).
\]

\item \textbf{Variogram score (VS).}
For exponent \(p\in(0,2]\) and nonnegative weights \(w^{k\ell}\) on component pairs \((k<\ell)\),
the Variogram Score at \((t,i)\) is
\[
\mathrm{VS}_{p,t}(i)
=\sum_{k<\ell} w^{k\ell}\left(\,|\tilde{y}_t^k(i)-\tilde{y}_t^\ell(i)|^{p}
-\mathbb{E}|\mathbf{X}^k-\mathbf{X}^\ell|^{p}\,\right)^{2}.
\]
 Ensemble estimator:
\[
\widehat{\mathrm{VS}}_{p,t}(i)
=\sum_{k<\ell}w^{k\ell}\left(\,|\tilde{y}_t^k(i)-\tilde{y}_t^\ell(i)|^{p}
-\frac{1}{M}\sum_{m=1}^{M}\big|\tilde{x}_{t,m}^k(i)-\tilde{x}_{t,m}^\ell(i)\big|^{p}\,\right)^{2}.
\]
Time-averaged per grid cell:
\[
\mathrm{VS}_{p}(i)=\frac{1}{T}\sum_{t=1}^{T}\widehat{\mathrm{VS}}_{p,t}(i).
\]
This study uses \(p=0.5\) and uniform weights (\(w^{k\ell}=1\)) following the setup of previous studies \citep{EsquivelGonzalez2026climdit}
\end{itemize}


\bibliographystyle{ametsocV6}
\bibliography{references}

\end{document}